\documentclass[10pt,journal]{IEEEtran}
\usepackage{amsmath,amssymb,amsfonts,mathrsfs,epsfig}
\usepackage{algorithmic}
\usepackage{algorithm}
\usepackage{array}
\usepackage[caption=false,font=normalsize,labelfont=sf,textfont=sf]{subfig}
\usepackage{textcomp}
\usepackage{stfloats}
\usepackage{url}
\usepackage{verbatim}
\usepackage{graphicx}
\usepackage{cite}
\newtheorem{remark}{Remark}

\begin{document}

\title{Flexible Intelligent Layered Metasurfaces for \\ Downlink Multi-user MISO Communications}

\author{Hong Niu,~\IEEEmembership{}
        Jianchang An,~\IEEEmembership{}
        and Chau Yuen,~\IEEEmembership{Fellow, IEEE}


\thanks{

H. Niu, J. An, and C. Yuen are with the School of Electrical and Electronics Engineering, Nanyang Technological University, Singapore 639798 (E-mail: \{hong.niu, jiancheng.an, chau.yuen\}@ntu.edu.sg).

}
}


\markboth{IEEE}%
{Shell \MakeLowercase{\textit{et al.}}: }
\maketitle

\begin{abstract}
Stacked intelligent metasurfaces (SIMs) have recently gained attention as a paradigm for wave-domain signal processing with reduced reliance on costly radio-frequency (RF) chains. However, conventional SIMs rely on uniform inter-layer spacing and require deep stacking to ensure processing capability, resulting in severe power attenuation in practice. To address this issue, we propose a flexible intelligent layered metasurface (FILM) architecture consisting of two shape-controllable flexible metasurface layers. By replacing rigid metasurfaces with flexible ones in both layers, the transmission coefficient matrix can be dynamically adjusted, significantly decreasing the number of required layers while maintaining signal processing performance. Firstly, we develop a two-layer FILM-assisted multi-user multiple-input single-output (MU-MISO) system, wherein we formulate a channel fitting problem aimed at reducing the difference between the FILM-induced and target channels. Then, we solve this non-convex problem by employing an alternating optimization (AO) method, featuring closed-form phase shift updates and a gradient descent-based shape optimization. Furthermore, we analyze the upper bound on sum-rate and the complexity of computation to provide insights into design trade-offs. Finally, simulation results demonstrated that the proposed transmissive FILM architecture achieves over 200\% improvement in sum-rate and more than 7 dB bit-error rate (BER) gain compared to the conventional seven-layer SIMs.
\end{abstract}

\begin{IEEEkeywords}
Stacked intelligent metasurfaces (SIMs), multi-user multiple-input single-output (MU-MISO), flexible intelligent metasurfaces (FIMs), power efficiency, wave-based beamforming.
\end{IEEEkeywords}

\IEEEpeerreviewmaketitle

\vspace*{-3mm}
\section{Introduction}

\IEEEPARstart{S}{tacked} intelligent metasurfaces (SIMs) have recently emerged as a novel paradigm in the field of electromagnetic (EM) wave manipulation and wireless communication \cite{SIM01}-\cite{SIM05}. Unlike conventional reconfigurable intelligent surfaces (RISs) that operate in a single layer \cite{RIS01}-\cite{RIS10}, SIMs leverage a multi-layered architecture, where each layer is composed of tunable meta-atoms capable of independently controlling the phase shift of incident waves \cite{SIM06}-\cite{SIM10}. By stacking multiple such layers with precisely designed inter-layer spacing, SIMs can achieve more complex and high-dimensional transformations of wavefronts, such as beamforming \cite{beam1}-\cite{beam3}, focusing \cite{focus1,focus2}, and multi-user (MU) channel customization \cite{MU1,MU2}. This layered configuration significantly enhances the degrees of freedom (DoFs) in wave-domain signal processing, offering a hardware-efficient alternative to conventional costly radio-frequency (RF) components. SIMs are particularly attractive for applications in holographic multiple-input multiple-output (MIMO) \cite{MIMO1}-\cite{MIMO3}, MU, and secure wireless communications \cite{secure1}-\cite{secure4}, where both performance and hardware cost are critical considerations.

To enable diverse signal processing functions such as image classification, \emph{Liu et al.} first proposed a programmable diffractive deep neural network architecture utilizing multi-layer metasurfaces \cite{Liu1}. Building upon this foundation, \emph{An et al.} integrated SIMs with MIMO transceivers to realize tasks including beamforming and combining \cite{An1}. The cascaded multi-layer configuration of SIMs enhances spatial DoFs while alleviating reliance on expensive RF chains.

Despite their promising potential in improving wireless communication, SIMs face two critical challenges stemming from their multi-layered design. First, the increased number of layers leads to a proliferation of meta-atoms, complicating system optimization. Second, signal transmission across multiple layers causes inevitable power attenuation, adversely affecting power efficiency \cite{Att1}. To tackle the issue of excessive layers, \emph{Niu et al.} proposed a meta-fiber-connected two-layer SIM architecture, demonstrating superior performance compared to conventional seven-layer designs \cite{fiber1}. Nevertheless, the cost and complexity of meta-fiber deployment pose practical barriers. Thus, devising low-cost approaches to reduce the number of SIM layers remains an open challenge.

Fundamentally, the multi-layer requirement of SIMs arises from their fixed transmission coefficient matrices. By cascading multiple fixed matrices, SIMs achieve the desired output signals with arbitrary amplitude and phase through meta-atom phase adjustments. This insight prompts a critical question: \textit{Can the number of layers be substantially reduced by employing a cost-effective, tunable transmission coefficient matrix?} Flexible intelligent metasurfaces (FIMs), a recently introduced technology, present a promising approach to address this challenge \cite{FIM1}.

Enabled by advances in micro/nanofabrication, FIMs are fabricated on deformable substrates such as polydimethylsiloxane (PDMS), enabling concurrent electromagnetic wave manipulation and physical surface reconfiguration. While existing work has explored FIMs for boosting wireless communication performance, their potential in signal processing applications remains largely untapped. Leveraging their shape-morphing capability, FIMs provide additional DoFs that enhance directional control and sensing accuracy \cite{FIM2}-\cite{FIM5}. Compared with fluid antenna systems (FAS) and movable antenna systems (MAS) \cite{FAS1}-\cite{MAS1}, FIMs offer faster, continuous surface adjustment and support compact, conformal deployment on diverse platforms.

\begin{figure*}[!htp]
\centering
\includegraphics[width=1\textwidth]{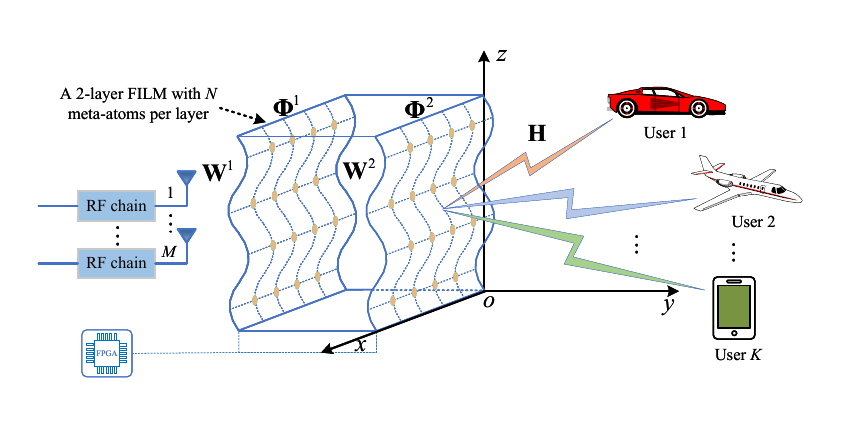}
\caption{Framework of FILM-assisted MU-MISO wireless communications.}
\label{fig_1}
\end{figure*}

Motivated by these developments, we propose a novel two-layer flexible intelligent layered metasurface (FILM) architecture that replaces the rigid meta-atoms in conventional SIMs with flexible ones on both layers. This design delivers two key benefits. First, adjusting meta-atom positions allows fine-tuning of the transmission coefficient matrix, substantially reducing the number of layers required in conventional SIMs. Second, although minor positional changes in FIMs yield limited gains in conventional wireless communication, mainly in highly scattering environments, the impact on signal processing tasks is profound, enabling more precise waveform customization. Hence, the proposed FILM architecture effectively integrates the strengths of both SIM and FIM, while mitigating their respective limitations. The key contributions of this work can be outlined as follows.
\begin{enumerate}

  \item \textbf{FILM Architecture:} We propose a two-layer FILM-assisted MU multiple-input single-output (MISO) system, enabling wave-based beamforming as illustrated in Fig. \ref{fig_1}. The integration of FILM at the transmitter allows direct transmission of multiple data streams from RF chains to users, while reducing the required number of layers compared to conventional SIM designs.

  \item \textbf{Phase-Shift and Shape Optimizations:} We formulate the channel fitting problem as a joint optimization over the FILM's phase shifts and shape. To address this non-convexity problem, we propose an alternating optimization (AO) algorithm that iteratively updates the phase shift using a closed-form solution and refines the shape through gradient descent (GD).

  \item \textbf{Theoretical Analysis:} An upper bound on sum-rate is obtained to offer analytical insight into the FILM design. In addition, computational complexity analysis reveals that the proposed AO algorithm exhibits cubic scaling with respect to the number of meta-atoms in each layer.

  \item \textbf{Performance Evaluation:} Simulation results reveal that the proposed two-layer FILM strikes an effective balance between SIM and FIM in channel fitting performance, while significantly outperforming both in throughput. Specifically, FILM achieves over 200\% improvement in sum-rate and more than 7 dB bit-error rate (BER) gain compared to the conventional seven-layer SIM architecture. Moreover, the single-layer FIM structure may not achieve accurate channel fitting, resulting in a BER that remains at a plateau near ${10^{ - 2}}$. In contrast, the proposed FILM reduces the BER to below ${10^{ - 5}}$.

\end{enumerate}

The remainder of this paper is organized as follows. Section II introduces the system model of the FILM-assisted MU-MISO system and formulates the corresponding channel fitting problem. Section III presents the joint optimization framework for the phase shifts and shapes of the metasurfaces in the FILM. In Section IV, the sum-rate upper bound and the computational complexity of the proposed algorithm are analyzed. Section V presents simulation outcomes and performance comparisons, followed by the conclusions in Section VI.

\textit{Notation}: In the following, lowercase bold letters and uppercase bold letters indicate vectors and matrices, respectively. $\left|  \cdot  \right|$, $\left\|  \cdot  \right\|$, and  ${\left(  \cdot  \right)^H}$ stand for the modulus, norm, and conjugate transpose operations, respectively. $\mathbb{E}\left(  \cdot  \right)$ and $\log \left(  \cdot  \right)$ represent the expectation operation and the logarithmic function with a cardinality of 2. ${X_{i,j}}$ and ${x_i}$ denote the element in $i$-th row and $j$-th column of the matrix ${\mathbf{X}}$ and the $i$-th element of the vector ${\mathbf{x}}$, respectively. ${\text{diag}}\left( {\mathbf{v}} \right)$ and $\arctan \left( x \right) = {\tan ^{ - 1}}\left( x \right)$ refer to a diagonal matrix with a vector ${\mathbf{v}}$ and an arc-tangent function, respectively. $j = \sqrt { - 1} $ and $\left\lceil x \right\rceil $ indicate the imaginary unit and the nearest integer no less than $x$, respectively. ${\mathbb{C}^{m \times n}}$ and $\mathcal{O}\left( {{f_N}} \right)$ denote complex matrices in $m \times n$ spaces and asymptotic time complexity proportional to ${{f_N}}$, respectively. $\,\bmod \,\left( { \cdot , \cdot } \right)$, $\sum$ and $\partial f/\partial x$ represent the operations of returning remainder, summing, and taking partial derivatives, respectively. ${{\mathbf{I}}_N}$ denotes an $N$-order unit matrix. $ {\cal C}{\cal N}\left( {u,{\sigma ^2}} \right)$ stands for a complex Gaussian distribution characterized by mean ${\mathbf{\mu }}$ and variance ${{\sigma ^2}}$.



\section{System Model}



In this section, we present the structure of a two-layer FILM. The corresponding channel models are then introduced, followed by the formulation of the optimization problem.

\subsection{FILM Model}

As observed from Fig. \ref{fig_1}, we consider an FILM-assisted MU-MISO system, where an FILM is deployed at the base station (BS) to perform beamforming towards $K$ users. Specifically, a two-layer FILM is assumed, with each layer comprising $N = {N_x} \times {N_z}$ meta-atoms, where ${N_x}$ and ${N_z}$ denote the number of meta-atoms along the $x$- and $z$-axes, respectively. The meta-atoms on each layer are arranged as a flexible uniform planar array (UPA) lying in the $x$-$o$-$z$ plane. Unlike conventional fixed arrays, the FILM is capable of dynamically reshape its surface. This is achieved by applying a magnetic field parallel to the surface plane, which generates a Lorentz force along the surface normal (i.e., the $y$-axis), allowing each meta-atom to be independently displaced in that direction.

Moreover, the $n$-th meta-atom on the $l$-th layer can continuously adjust the phase shift of the incident EM signal within the range of $\theta _n^l \in \left[ {0,2\pi } \right),n = 1,2, \ldots ,N,l = 1,2$. Accordingly, its transmission coefficient    is given by $\phi _n^l = {e^{j\theta _n^l}}$, and the overall transmission matrix of the $l$-th layer is written as ${{\mathbf{\Phi }}^l} = {\text{diag}}\left( {{{\left[ {\phi _1^l,\phi _2^l, \cdots ,\phi _N^l} \right]}^T}} \right)$.

Furthermore, the transmission matrix between the BS and the first layer of FILM is denoted by ${{\mathbf{W}}^{\left[ 1 \right]}} \in {\mathbb{C}^{N \times M}}$, where $M \geqslant K$ represents the number of RF chains at the BS. Similarly, the transmission matrix between the first and second layers of FILM is denoted by ${{\mathbf{W}}^{\left[ 2 \right]}} \in {\mathbb{C}^{N \times N}}$. By treating the BS as the 0-th layer, we denote the transmission matrix linking the $\left( {l - 1} \right)$-th and $l$-th layers as ${{\mathbf{W}}^{\left[ l \right]}}$. Each element $W_{n,\tilde n}^{\left[ l \right]}$ in the $n$-th row and $\tilde n$-th-th column represents the diffraction coefficient between the $\tilde n$-th meta-atom on the $\left( {l - 1} \right)$-th layer and the $n$-th meta-atom on the $l$-th layer, given by\footnote{Brackets are used in the superscripts to prevent misinterpretation of the layer indices as power terms in subsequent derivations.}
\begin{equation}\label{wnn:1}
W_{n,\tilde n}^{\left[ l \right]} = \frac{{{A_t}\cos \chi _{n,\tilde n}^{\left[ l \right]}}}{{d_{n,\tilde n}^{\left[ l \right]}}}\left( {\frac{1}{{2\pi d_{n,\tilde n}^{\left[ l \right]}}} - j\frac{1}{\lambda }} \right){e^{j2\pi d_{n,\tilde n}^{\left[ l \right]}/\lambda }}, l=1,2,
\end{equation}
where ${{A_t}}$ denotes the area occupied by each meta-atom, ${\chi _{n,\tilde n}^{\left[ l \right]}}$ is the included angle between the normal of the ($l-1$)-th layer and the EM wave propagation direction, $\lambda $ represents the wavelength, and ${d_{n,\tilde n}^{\left[ l \right]}}$ stands for the transmission distance. Equation (\ref{wnn:1}) adopts a near-field forward propagation model between adjacent metasurfaces, with the diffraction coefficient characterized by the Rayleigh-Sommerfeld diffraction theory \cite{w1}.

Consequently, the corresponding transmission distance between the $\tilde n$-th meta-atom on the ($l-1$)-th layer of FILM and the $n$-th meta-atom on its $l$-th layer is shown as
\begin{equation}\label{rnn:1}
d_{n,\tilde n}^{\left[ l \right]} = \left\| {{\mathbf{p}}_n^{\left[ l \right]} - {\mathbf{p}}_{\tilde n}^{\left[ {l - 1} \right]}} \right\|,
\end{equation}
where ${{\mathbf{p}}_n^{\left[ l \right]}}$ denotes the coordinate of the $n$-th meta-atom on the $l$-th layer, expressed as
\begin{equation}\label{rnn:2}
{\mathbf{p}}_n^{\left[ l \right]} = \left[ {\begin{array}{*{20}{c}}
  {x_n^{\left[ l \right]}}&{y_n^{\left[ l \right]}}&{z_n^{\left[ l \right]}}
\end{array}} \right],
\end{equation}
with ${x_n^{\left[ l \right]}}$, ${y_n^{\left[ l \right]}}$, and ${z_n^{\left[ l \right]}}$ representing the spatial coordinates along the $x$-, $y$-, $z$-axes, respectively. Since Lorentz force is applied only along the $y$-axis \cite{FIM4}, ${x_n^{\left[ l \right]}}$ and ${z_n^{\left[ l \right]}}$ remain fixed, while ${y_n^{\left[ l \right]}}$ can vary within the range
\begin{equation}
{{\bar y}^{\left[ l \right]}} - \left| {{y_{\max }}} \right| \leqslant y_n^{\left[ l \right]} \leqslant {{\bar y}^{\left[ l \right]}} + \left| {{y_{\max }}} \right|,
\end{equation}
where ${{\bar y}^{\left[ l \right]}}$ denotes the $y$-coordinate of the center position of the $l$-th metasurface, and $\left| {{y_{\max }}} \right|$ represents the maximum morphing range for leftward and rightward displacements.

Building upon the above model, the beamforming matrix of FILM is formulated as
\begin{equation}\label{PRE:1}
{\mathbf{P}} = {{\mathbf{\Phi }}^2}{{\mathbf{W}}^{\left[ 2 \right]}}{{\mathbf{\Phi }}^1}{{\mathbf{W}}^{\left[ 1 \right]}} \in {\mathbb{C}^{N \times K}}.
\end{equation}

\subsection{Channel Model}

We consider the scenario with $K$ users, where the $k$-th user is located at an elevation angle of ${{\theta _k} \in \left[ {0,\pi } \right]}$ and an azimuth angle of ${{\varphi _k} \in \left[ {0,\pi } \right]}$. For simplicity, the coordinate axis is aligned such that ${{\bar y}^{\left[ 2 \right]}}=0$, as illustrated in Fig. \ref{fig_1}. The transmitting steering vector from the second layer of FILM to the $k$-th user can be expressed as
\begin{equation}\label{rnn:3}
{\mathbf{a}}\left( {{\theta _k},{\varphi _k},{{\mathbf{y}}^{\left[ 2 \right]}}} \right) = \left[ {{{\mathbf{a}}_z}\left( {{\theta _k}} \right) \otimes {{\mathbf{a}}_x}\left( {{\theta _k},{\varphi _k}} \right)} \right] \odot {{\mathbf{a}}_y}\left( {{\theta _k},{\varphi _k},{{\mathbf{y}}^{\left[ 2 \right]}}} \right),
\end{equation}
for $k = 1,2, \ldots ,K$, where ${{{\mathbf{a}}_z}\left( {{\theta _k}} \right)}$ and ${{{\mathbf{a}}_x}\left( {{\theta _k},{\varphi _k}} \right)}$ denote the array response vectors along the $z$- and $x$-axes, respectively, defined as
\begin{equation}\label{rn1:1}
{{\mathbf{a}}_x}\left( {{\theta _k},{\varphi _k}} \right) \triangleq {\left[ {1,{e^{ - j{k_c}{v_{x,k}}}}, \cdots ,{e^{ - j{k_c}\left( {{N_x} - 1} \right){v_{x,k}}}}} \right]^T},
\end{equation}
\begin{equation}
{{\mathbf{a}}_z}\left( {{\theta _k}} \right) \triangleq {\left[ {1,{e^{ - j{k_c}{v_{z,k}}}}, \cdots ,{e^{ - j{k_c}\left( {{N_z} - 1} \right){v_{z,k}}}}} \right]^T},
\end{equation}
with ${v_{x,k}} = {d_x}\sin {\theta _k}\cos {\varphi _k}$ and ${v_{z,k}} = {d_z}\cos {\theta _k}$. Here, ${d_x}$ and ${d_z}$ represent the spacing between adjacent meta-atoms of the FILM along the $x$- and $z$-axes, respectively, and ${k_c} = 2\pi /\lambda $ is the wavenumber associated with the wavelength $\lambda$. Let ${{\mathbf{y}}^{\left[ 2 \right]}} = \left[ {y_1^{\left[ 2 \right]},y_2^{\left[ 2 \right]}, \cdots ,y_N^{\left[ 2 \right]}} \right] \in {\mathbb{R}^N}$ denote the $y$-coordinates of the meta-atoms on the second layer of FILM. The array response vectors along the $y$-axis, ${{\mathbf{a}}_y}\left( {{\theta _k},{\varphi _k},{{\mathbf{y}}^{\left[ 2 \right]}}} \right)$, can then by expressed as
\begin{equation}
\begin{gathered}
  {{\mathbf{a}}_y}\left( {{\theta _k},{\varphi _k},{{\mathbf{y}}^{\left[ 2 \right]}}} \right) = \left[ {{e^{ - j{k_c}y_1^{\left[ 2 \right]}{v_{y,k}}}},} \right.{e^{ - j{k_c}y_2^{\left[ 2 \right]}{v_{y,k}}}}, \hfill \\
  \ \ \ \ \ \ \ \ \ \ \ \ \ \ \ \ \ \ \ \ {\left. { \cdots ,{e^{ - j{k_c}y_N^{\left[ 2 \right]}{v_{y,k}}}}} \right]^T} \in {\mathbb{C}^N}, \hfill \\
\end{gathered}
\end{equation}
with ${{v_{y,k}} = \sin {\theta _k}\sin {\varphi _k}}$.

By stacking the transmitting steering vectors of all users, the array's response matrix can be obtained as
\begin{equation}\label{AR:1}
{\mathbf{A}} \triangleq \left[ {{\mathbf{a}}\left( {{\theta _1},{\varphi _1},{{\mathbf{y}}^{\left[ 2 \right]}}} \right),{\mathbf{a}}\left( {{\theta _2},{\varphi _2},{{\mathbf{y}}^{\left[ 2 \right]}}} \right), \cdots ,{\mathbf{a}}\left( {{\theta _K},{\varphi _K},{{\mathbf{y}}^{\left[ 2 \right]}}} \right)} \right].
\end{equation}

Accordingly, the channel from the second layer of FILM to the users can be given by
\begin{equation}\label{P:1}
{\mathbf{H}} = {\text{diag}}\left( {{\beta _1},{\beta _2}, \cdots ,{\beta _K}} \right){{\mathbf{A}}^H}  \in {\mathbb{C}^{K \times N}},
\end{equation}
where ${\boldsymbol{\beta }} = \left[ {{\beta _1},{\beta _2}, \cdots ,{\beta _K}} \right]$ is composed of distance-dependent path-loss coefficients ${\beta _k^2} = - 20{\log _{10}}\left( {4\pi /\lambda } \right) - 10b{\log _{10}}\left( {{d_k}} \right),i = 1,2, \cdots ,K,$ with $b$ denoting the path loss exponent and ${{d_k}}$ representing the distances from the second layer of FILM to the $k$-th user \cite{An1}.

\subsection{Problem Formulation}

Through the beamforming of FILM, channel diagonalization facilitates the simultaneous transmission and retrieval of multiple data streams between the BS and MUs. Specifically, tuning the phase shifts and shape of the FILM transforms the end-to-end channel matrix into a desired identity matrix as
\begin{equation}\label{UM:1}
{\mathbf{HP}} \to {\mathbf{I}}_K.
\end{equation}

Based on (\ref{UM:1}), the optimization problem can be formulated as
\begin{subequations}\label{pf:1}
\begin{align}
\mathop {\min }\limits_{\theta _n^l,y_n^{\left[ l \right]},\alpha }  &J = \left\| {{\mathbf{HP}} - \alpha {{\mathbf{I}}_K}} \right\|_F^2, \label{Za}\\
{\text{s}}{\text{.t}}{\text{.}} \;\;\; &{\mathbf{P}} = {{\mathbf{\Phi }}^2}{{\mathbf{W}}^{\left[ 2 \right]}}{{\mathbf{\Phi }}^1}{{\mathbf{W}}^{\left[ 1 \right]}}, \label{Zb}\\
&{\mathbf{H}} = {\text{diag}}\left( {{\beta _1},{\beta _2}, \cdots ,{\beta _K}} \right){{\mathbf{A}}^H},\label{Zc}\\
&{{\mathbf{\Phi }}^l} = {\text{diag}}\left( {{{\left[ {{e^{j\theta _1^l}},{e^{j\theta _2^l}}, \cdots ,{e^{j\theta _N^l}}} \right]}^T}} \right),\label{Zd}\\
&\left| {y_n^{\left[ l \right]} - {{\bar y}^{\left[ l \right]}}} \right| \leqslant \left| {{y_{\max }}} \right|, l=1,2,\label{Ze}\\
&\alpha  \in {\mathbb{R}^*},\label{Zf}
\end{align}
\end{subequations}
where $\alpha $ denotes the channel gain, characterizing the quality of the effective channel. The objective function in (\ref{Za}) seeks to realize an identity channel matrix through the beamforming at FILM, as described in (\ref{UM:1}). The constraints in (\ref{Zb}) and (\ref{Zc}) specify the end-to-end channel from the BS to the MUs, as formulated in (\ref{PRE:1}) and (\ref{P:1}). In addition, the constraint in (\ref{Zd}) reflects the impact of meta-atoms is predominantly on the phase shift rather than the amplitude. Lastly, the constraint in (\ref{Ze}) restricts the maximum morphing range, imposed by the elastic properties of the surface material.

Owing to the non-convex constants in (\ref{Zd}) and (\ref{Ze}), along with the strongly coupled variables in (\ref{Za}), obtaining the optimal solution to Problem (\ref{pf:1}) is quite challenging. Although prior works have explored the phase shift design in (\ref{Zd}), the shape design in (\ref{Ze}) has yet to be fully addressed. To address this gap, we derive closed-form expressions for phase shifts and propose a GD algorithm for morphing shapes of the two metasurfaces. The closed-form solutions provide intuitive insight into the phase shfit design and offers a lower computational complexity in contrast to conventional optimization methods. Meanwhile, the proposed GD algorithm captures the impact of FILM's shape on both the internal near-field and external far-field channels, achieving the desired channel fitting effect by jointly adjusting the channels at both ends.

\begin{remark}
As indicated in (\ref{UM:1}), multiple data streams are directly conveyed from the BS to the MUs, which substantially reduces both transmission and reception complexity. In contrast to \cite{An1}, the proposed scheme additionally balances the quality of each data stream, thereby reducing the likelihood of unfavorable channel conditions.
\end{remark}

\section{Joint Design of FILM Phase Shifts,\\ Surface Shape, and Channel Gains}

In this section, we propose an AO algorithm that jointly designs the FILM shape, FILM phase shifts, and equivalent channel gain. Specifically, we derived closed-form expressions for the phase shifts and channel gain, and further propose a GD algorithm for shape optimization.

\subsection{Closed-form Solution for Phase Shifts}

Here, a closed-form solution for each phase shift is derived by evaluating its partial derivative and solving for its stationary points. Thus, Problem (\ref{pf:1}) with respect to the phase shifts can be expressed as
\begin{subequations}\label{pf:2}
\begin{align}
\mathop {\min }\limits_{\phi _n^l} \;&J = {\left\| {{\mathbf{A}^l}{{\mathbf{\Phi }}^l}{\mathbf{B}^l} - {\mathbf{I}}} \right\|^2}, \label{Za2}\\
{\text{s}}{\text{.t}}{\text{.}} \;\;&\text{(\ref{Zb}), (\ref{Zd})}.
\end{align}
\end{subequations}
where ${{\mathbf{I}} = \alpha {{\mathbf{I}}_K}}$ denotes the desired identity channel matrix. ${\mathbf{A}^l} \in {\mathbb{C}^{K \times N}}$ represents the equivalent channel from the $l$-th FILM layer to the MUs, while ${\mathbf{B}^l} \in {\mathbb{C}^{N \times K}}$ stands for the equivalent channel from the BS to $l$-th FILM layer (excluding meta-atoms on the $l$-th layer), formulated by
\begin{equation}
{{\mathbf{A}}^1} = {\mathbf{H}}{{\mathbf{\Phi }}^2}{{\mathbf{W}}^{\left[ 2 \right]}},{{\mathbf{B}}^1} = {{\mathbf{W}}^{\left[ 1 \right]}},{{\mathbf{A}}^2} = {\mathbf{H}},{{\mathbf{B}}^2} = {{\mathbf{W}}^{\left[ 2 \right]}}{{\mathbf{\Phi }}^1}{{\mathbf{W}}^{\left[ 1 \right]}}.
\end{equation}

By definition, the function in (\ref{Za2}) is rewritten as
\begin{equation}\label{Re:2}
J = \sum\limits_{k = 1}^K {\sum\limits_{j = 1}^K {{{\left| {\sum\limits_{i = 1}^N {A_{k,i}^lB_{i,j}^l{e^{j\theta _i^l}}}  - {I_{k,j}}} \right|}^2}} } ,
\end{equation}
where ${A_{k,i}^l}$ denotes the element in the $k$-th row and $i$-th column of the matrix ${{\mathbf{A}}^l}$, $B_{i,j}^l$ represents the element in the $i$-th row and $j$-th column of the matrix ${{\mathbf{B}}^l}$, and ${I_{k,j}}$ stands for the element in the $k$-th row and $j$-th column of the matrix ${{\mathbf{I}}}$.

To simplify the expression in (\ref{Re:2}), we defined the tensors ${{\mathbf{L}}^l} \in {\mathbb{C}^{K \times N \times K}}$ as
\begin{equation}
{L_{k,i,j}^l = A_{k,i}^lB_{i,j}^l}.
\end{equation}

Accordingly, the expression in (\ref{Re:2}) can be simplified as
\begin{equation}\label{Re:3}
J = \sum\limits_{k = 1}^K {\sum\limits_{j = 1}^K {{{\left| {\sum\limits_{i = 1}^{N} {\left| {L_{k,i,j}^l} \right|{e^{j\left( {\xi _{k,i,j}^l + \theta _i^l} \right)}}}  - \left| {{I_{k,j}}} \right|{e^{j{\zeta _{k,j}}}}} \right|}^2}} } ,
\end{equation}
where ${\left| {L_{k,i,j}^l} \right|}$ and ${\xi _{k,i,j}^l}$ denote the amplitude and phase of ${L_{k,i,j}^l}$, i.e., $L_{k,i,j}^l = \left| {L_{k,i,j}^l} \right|{e^{j\xi _{k,i,j}^l}}$, while ${\left| {{I_{k,j}}} \right|}$ and ${{\zeta _{k,j}}}$ denote those of ${{I_{k,j}}}$, i.e., ${{I_{k,j}} = \left| {{I_{k,j}}} \right|{e^{j{\zeta _{k,j}}}}}$.

In addition, (\ref{Re:3}) can be expanded as
\begin{equation}\label{Re:4}
\begin{gathered}
  J = \sum\limits_{k = 1}^K {\sum\limits_{j = 1}^K {\left( {{{\left| {{I_{k,j}}} \right|}^2} + \sum\limits_{i = 1}^{N} {{{\left| {L_{k,i,j}^l} \right|}^2}} } \right)} }  \hfill \\
   + 2\sum\limits_{k = 1}^K {\sum\limits_{j = 1}^K {\sum\limits_{q = i + 1}^{N} {\sum\limits_{i = 1}^{N - 1} {\left| {L_{k,i,j}^l} \right|\left| {L_{k,q,j}^l} \right|} } } }  \hfill \\
  \cos \left( {\xi _{k,i,j}^l + \theta _i^l - \xi _{k,q,j}^l - \theta _q^l} \right) \hfill \\
   - 2\sum\limits_{k = 1}^K {\sum\limits_{j = 1}^K {\sum\limits_{i = 1}^{N} {\left| {L_{k,i,j}^l} \right|\left| {{I_{k,j}}} \right|} \cos \left( {\xi _{k,i,j}^l + \theta _i^l - {\zeta _{k,j}}} \right)} } . \hfill \\
\end{gathered}
\end{equation}

To obtain the optimal value of the $n$-th phase shift on the $l$-th layer $\theta _n^l$, we calculate the partial derivative of the objective $J$ in (\ref{Re:4}) with respect to $\theta _n^l$. It is worth noting that only the second and third terms in (\ref{Re:4}) are functions of $\theta_n^l$. Specifically, the second term must be evaluated differently depending on whether the indices $q$ and $i$ match $n$. Consequently, the derivative of $J$ with respect to $\theta_n^l$ can be expressed as
\begin{equation}\label{pd:1}
\frac{{\partial J}}{{\partial \theta _n^l}} = 2\left( {e_n^l\cos \theta _n^l - f_n^l\sin \theta _n^l - g_n^l\cos \theta _n^l - h_n^l\sin \theta _n^l} \right),
\end{equation}
where $e_n^l$, $f_n^l$, $g_n^l$, and $h_n^l$ are defined as
\begin{subequations}
\begin{align}
  &e_n^l = \sum_{k = 1}^K \sum_{j = 1}^K
        \sum_{\substack{i = 1, \\ i \ne n}}^N
        \left| L_{k,i,j}^l \right|
        \left| L_{k,n,j}^l \right|
        \sin \!\left( \xi_{k,i,j}^l + \theta_i^l - \xi_{k,n,j}^l \right),    \hfill  \\
  &f_n^l = \sum\limits_{k = 1}^K {\sum\limits_{j = 1}^K {\left| {L_{k,n,j}^l} \right|\left| {{I_{k,j}}} \right|\sin \left( {\xi _{k,n,j}^l - {\zeta _{k,j}}} \right)} ,}     \\
  &g_n^l = \sum_{k = 1}^K \sum_{j = 1}^K
        \sum_{\substack{i = 1, \\ i \ne n}}^N
        \left| L_{k,i,j}^l \right|
        \left| L_{k,n,j}^l \right|
        \cos \!\left( \xi_{k,i,j}^l + \theta_i^l - \xi_{k,n,j}^l \right),      \\
  &h_n^l = \sum\limits_{k = 1}^K {\sum\limits_{j = 1}^K {\left| {L_{k,n,j}^l} \right|\left| {{I_{k,j}}} \right|\cos \left( {\xi _{k,n,j}^l - {\zeta _{k,j}}} \right)} .}  \\
\end{align}
\end{subequations}

Setting the partial derivative $\partial J/\partial \theta_n^l$ to zero yields two possible solutions for $\theta_n^l$ within the interval $\left[0,2\pi\right)$ as
\begin{subequations}\label{ps:1}
\begin{align}
 & \theta _{n,1}^l = \arctan \left( {\frac{{{{e_n^l}} + {{f_n^l}}}}{{{{g_n^l}} - {{h_n^l}}}}} \right), \label{t1:1}\\
 & \theta _{n,2}^l = \arctan \left( {\frac{{{{e_n^l}} + {{f_n^l}}}}{{{{g_n^l}} - {{h_n^l}}}}} \right) + \pi . \label{t1:2}
\end{align}
\end{subequations}

The optimal $\theta_n^l$ is chosen from $\theta_{n,1}^l$ and $\theta_{n,2}^l$ based on which yields a smaller value of $J$.

\subsection{GD Algorithm for Shape}

Here, we propose a GD algorithm for shape optimization by computing the gradient of the objective function concerning the $y$-coordinate of each meta-atom. Specifically, the FLIM shape optimization can be expressed as
\begin{subequations}\label{pf:3}
\begin{align}
\mathop {\min }\limits_{y_n^{\left[ l \right]}} \;&J = \left\| {{\mathbf{H}}{{\mathbf{\Phi }}^2}{{\mathbf{W}}^{\left[ 2 \right]}}{{\mathbf{\Phi }}^1}{{\mathbf{W}}^{\left[ 1 \right]}} - {\mathbf{I}}} \right\|_F^2, \label{Za3}\\
{\text{s}}{\text{.t}}{\text{.}} \;\;&\text{(\ref{Zb}), (\ref{Zc}), (\ref{Ze})}.
\end{align}
\end{subequations}

The shape design of the first and second FILM layers serves distinct purposes. The first layer controls the near-field channels from the BS to the first layer and from the first to the second layer, whereas the second layer governs both the near-field channel from the first to the second layer and the far-field channel from the second layer to the MUs. We discuss these two layers separately.

\subsubsection{The First FILM Layer Design}

Using the chain rule for gradient derivation, the gradient of $J$ concerning the $y$-coordinate of the $n$-th meta-atom on the first layer $y_n^{\left[ 1 \right]}$ is given by
\begin{equation}\label{fir:1}
\begin{gathered}
  \frac{{\partial J}}{{\partial y_n^{\left[ 1 \right]}}} = 2\Re \left\{ {{\text{Tr}}\left[ {{{\mathbf{C}}^H}\frac{{\partial {\mathbf{C}}}}{{\partial y_n^{\left[ 1 \right]}}}} \right]} \right\} \hfill \\
   = 2\Re \left\{ {{\text{Tr}}\left[ {{{\mathbf{C}}^H}{\mathbf{H}}{{\mathbf{\Phi }}^2}\left( {\frac{{\partial {{\mathbf{W}}^{\left[ 2 \right]}}}}{{\partial y_n^1}}{{\mathbf{\Phi }}^1}{{\mathbf{W}}^{\left[ 1 \right]}} + {{\mathbf{W}}^{\left[ 2 \right]}}{{\mathbf{\Phi }}^1}\frac{{\partial {{\mathbf{W}}^{\left[ 1 \right]}}}}{{\partial y_n^1}}} \right)} \right]} \right\}, \hfill \\
\end{gathered}
\end{equation}
where ${\mathbf{C}} = {\mathbf{HP}} - {\mathbf{I}}$ denotes the channel deviation between the actual and desired channels.

According to the expressions in (\ref{wnn:1}), we have
\begin{subequations}\label{fir:2}
\begin{align}
 & \frac{{\partial {{\mathbf{W}}^{\left[ 2 \right]}}}}{{\partial y_n^{\left[ 1 \right]}}} = \frac{{\partial {\mathbf{w}}_n^{\left[ 2 \right]}}}{{\partial y_n^{\left[ 1 \right]}}}{\mathbf{e}}_n^T,\\
 & \frac{{\partial {{\mathbf{W}}^{\left[ 1 \right]}}}}{{\partial y_n^{\left[ 1 \right]}}} = \frac{{\partial {{\left( {{\mathbf{w}}_n^{\left[ 1 \right]}} \right)}^H}}}{{\partial y_n^{\left[ 1 \right]}}}{{\mathbf{e}}_n},
\end{align}
\end{subequations}
where ${{\mathbf{e}}_n} = {\left( {0, \ldots ,0,1,0, \ldots ,0} \right)^T} \in {\mathbb{R}^N}$ denote the $n$-th standard basis vector, where the entry 1 is at the $n$-th position and all other entries are 0. Considering that the influence of a small position change on the signal amplitude is much smaller than that on the phase, the partial derivative can be approximated as
\begin{subequations}\label{fir:3}
\begin{align}
  {\left[ {\frac{{\partial {\mathbf{w}}_n^{\left[ 2 \right]}}}{{\partial y_n^{\left[ 1 \right]}}}} \right]_m} &\approx  - W_{m,n}^{\left[ 2 \right]}\frac{{y_m^{\left[ 2 \right]} - y_n^{\left[ 1 \right]}}}{{d_{m,n}^{\left[ 2 \right]}}}j2\pi /\lambda ,\\
  {\left[ {\frac{{\partial {{\left( {{\mathbf{w}}_n^{\left[ 1 \right]}} \right)}^H}}}{{\partial y_n^{\left[ 1 \right]}}}} \right]_k} & \approx W_{n,k}^{\left[ 1 \right]}\frac{{y_n^{\left[ 1 \right]} - y_k^{\left[ 0 \right]}}}{{d_{n,k}^{\left[ 1 \right]}}}j2\pi /\lambda ,
\end{align}
\end{subequations}
for $n = 1,2, \ldots ,N,m = 1,2, \cdots ,N$, and $k = 1,2, \cdots ,K$.

The partial derivative of the objective function $J$ with respect to $y_n^{\left[ 1 \right]}$ can be completed by substituting (\ref{fir:2}) and (\ref{fir:3}) into (\ref{fir:1}).

Based on (\ref{fir:1}), the shape of the first FILM layer design can be iteratively updated as
\begin{equation}
y_n^{\left[ 1 \right]} \leftarrow y_n^{\left[ 1 \right]} + \eta \frac{{\partial J}}{{\partial y_n^{\left[ 1 \right]}}},
\end{equation}
where $\eta  > 0$ denotes the step size at each iteration.

To enforce the maximum morphing range specified in (\ref{Ze}), a projection is employed to ensure that the positions of the meta-atoms on the first FILM layer remain within their allowable bounds, given by
\begin{equation}\label{lim:1}
y_n^{\left[ 1 \right]} = \min \left( {{{\bar y}^{[1]}} + \left| {{y_{\max }}} \right|,\max \left( {{{\bar y}^{[1]}} - \left| {{y_{\max }}} \right|,y_n^{\left[ 1 \right]}} \right)} \right).
\end{equation}

\begin{remark}
From (\ref{fir:1}), the shape design of the first FILM layer entails joint optimization of the near-field channels between the BS and the first layer, as well as between the first and second layers. Therefore, the proposed gradient descent algorithm is applicable to the middle layers of a multi-layer FILM (with more than two layers).
\end{remark}

\subsubsection{The Second FILM Layer Design}

Similarly, the gradient of $J$ concerning the $y$-coordinate of the $n$-th meta-atom on the second layer $y_n^{\left[ 2 \right]}$ is expressed as
\begin{equation}\label{sec:1}
\begin{gathered}
  \frac{{\partial J}}{{\partial y_n^{\left[ 2 \right]}}} = 2\Re \left\{ {{\text{Tr}}\left[ {{{\mathbf{C}}^H}\frac{{\partial {\mathbf{H}}{{\mathbf{\Phi }}^2}{{\mathbf{W}}^{\left[ 2 \right]}}}}{{\partial y_n^{\left[ 2 \right]}}}{{\mathbf{\Phi }}^1}{{\mathbf{W}}^{\left[ 1 \right]}}} \right]} \right\} \hfill \\
   = 2\Re \left\{ {{\text{Tr}}\left[ {{{\mathbf{C}}^H}\left( {\frac{{\partial {\mathbf{H}}}}{{\partial y_n^{\left[ 2 \right]}}}{{\mathbf{\Phi }}^2}{{\mathbf{W}}^{\left[ 2 \right]}} + {\mathbf{H}}{{\mathbf{\Phi }}^2}\frac{{\partial {{\mathbf{W}}^{\left[ 2 \right]}}}}{{\partial y_n^{\left[ 2 \right]}}}} \right){{\mathbf{\Phi }}^1}{{\mathbf{W}}^{\left[ 1 \right]}}} \right]} \right\}. \hfill \\
\end{gathered}
\end{equation}

According to the expressions in (\ref{wnn:1}), we have
\begin{equation}\label{sec:2}
\frac{{\partial {{\mathbf{W}}^{\left[ 2 \right]}}}}{{\partial y_n^{\left[ 2 \right]}}} = \frac{{\partial {{\left( {{\mathbf{w}}_n^{\left[ 2 \right]}} \right)}^H}}}{{\partial y_n^{\left[ 2 \right]}}}{{\mathbf{e}}_n},
\end{equation}
where the partial derivative can be approximated as
\begin{equation}
{\left[ {\frac{{\partial {{\left( {{\mathbf{w}}_n^{\left[ 2 \right]}} \right)}^H}}}{{\partial y_n^{\left[ 2 \right]}}}} \right]_m} \approx W_{n,m}^{\left[ 2 \right]}\frac{{y_n^{\left[ 2 \right]} - y_m^{\left[ 1 \right]}}}{{d_{n,m}^{\left[ 2 \right]}}}j2\pi /\lambda ,
\end{equation}
for $n = 1,2, \ldots,N$ and $m = 1,2, \cdots ,N$.

Based on (\ref{AR:1}) and (\ref{P:1}), we arrive at
\begin{equation}\label{sec:3}
\begin{gathered}
  \frac{{\partial {\mathbf{H}}}}{{\partial y_n^{\left[ 2 \right]}}} = {\text{diag}}\left( {\boldsymbol{\beta }} \right)\frac{{\partial {{\mathbf{A}}^H}}}{{\partial y_n^{\left[ 2 \right]}}} \hfill \\
   = {\text{diag}}\left( {\boldsymbol{\beta }} \right){\left[ {\begin{array}{*{20}{c}}
  {{\text{d}}y_{n,1}^{[2]}}&{{\text{d}}y_{n,2}^{[2]}}& \cdots &{{\text{d}}y_{n,K}^{[2]}}
\end{array}} \right]^H}{\mathbf{e}}_n^T, \hfill \\
\end{gathered}
\end{equation}
where ${\text{d}}y_{n,k}^{[2]},k=1,2, \cdots ,K$ are defined as
\begin{equation}
{\text{d}}y_{n,k}^{[2]} = j{k_c}{v_{y,k}}{e^{j{k_c}\left[ {\left( {{n_x} - 1} \right){v_{x,k}} + \left( {{n_z} - 1} \right){v_{z,k}} + y_n^{\left[ 2 \right]}{v_{y,k}}} \right]}}.
\end{equation}

The gradient of $J$ concerning $y_n^{\left[ 2 \right]}$ can be completed by substituting (\ref{sec:2}) and (\ref{sec:3}) into (\ref{sec:1}).

Based on (\ref{sec:1}), the shape of the second FILM layer design can be iteratively updated as
\begin{equation}
y_n^{\left[ 2 \right]} \leftarrow y_n^{\left[ 2 \right]} + \eta \frac{{\partial J}}{{\partial y_n^{\left[ 2 \right]}}},
\end{equation}
where $\eta  > 0$ denotes the step size at each iteration.

To enforce the maximum morphing range specified in (\ref{Ze}), a projection is employed to ensure that the positions of the meta-atoms on the first FILM layer remain within their allowable bounds, given by
\begin{equation}\label{lim:2}
y_n^{\left[ 2 \right]} = \min \left( {{{\bar y}^{[2]}} + \left| {{y_{\max }}} \right|,\max \left( {{{\bar y}^{[2]}} - \left| {{y_{\max }}} \right|,y_n^{\left[ 2 \right]}} \right)} \right).
\end{equation}

\begin{remark}
From (\ref{sec:1}), the shape design of the second FILM layer involves jointly optimizing the near-field channel between the first and second layers and the far-field channel from the second layer to the MUs. Hence, the proposed gradient descent algorithm can be applied to either the output layer of a multi-layer FILM (with more than two layers) or a single-layer FIM.
\end{remark}

\subsection{Closed-form Solution for Channel Gain}

In this subsection, we need to calculate the gradient of $J$ concerning $\alpha$, thus the objective function is rewritten as
\begin{equation}
J = \left\| {{\mathbf{G}} - \alpha {{\mathbf{I}}_K}} \right\|_F^2
\end{equation}
where ${\mathbf{G}} = {\mathbf{HP}}$ denotes the equivalent channel from the BS to the MUs.

Setting the gradient $\partial J/\partial \alpha $ to zero yields the unique solution as
\begin{equation}\label{alpha:1}
\alpha  = \frac{{{\text{tr}}\left( {{\mathbf{G}} + {{\mathbf{G}}^H}} \right)}}{{2K}}.
\end{equation}

By iteratively applying (\ref{fir:1}), (\ref{sec:1}), and (\ref{alpha:1}), the objective function $J$ gradually converges. The comprehensive steps of the AO algorithm are listed in Algorithm 1 for better understanding

\begin{algorithm} [h]
        \caption{The Proposed AO Algorithm for Solving (\ref{pf:1})}
		\label{alg_approx}
		\begin{algorithmic}[1]
        \REQUIRE
			$\theta _n^l,{\mathbf{p}}_n^{\left[ l \right]},\alpha $.
        \REPEAT
        \FOR {$l = 1;l \leqslant 2;l +  + $}
        \FOR {$n = 1;n \leqslant N;n +  + $}
           \STATE          Calculating $\theta _{n,1}^l$ and $\theta _{n,2}^l$ by applying (\ref{t1:1}) and (\ref{t1:2});
           \STATE          Updating $\theta _n^l$ with the one making $J$ smaller;
        \ENDFOR
        \ENDFOR
        \FOR {$n = 1;n \leqslant N;n +  + $}
           \STATE          Calculating $y_n^{\left[ 1 \right]}$ by applying (\ref{fir:1});
           \STATE          Limiting the range of $y_n^{\left[ 1 \right]}$ with (\ref{lim:1});
        \ENDFOR
        \FOR {$n = 1;n \leqslant N;n +  + $}
           \STATE          Calculating $y_n^{\left[ 2 \right]}$ by applying (\ref{sec:1});
           \STATE          Limiting the range of $y_n^{\left[ 2 \right]}$ with (\ref{lim:2});
        \ENDFOR
           \STATE          Updating $\alpha$ by applying (\ref{alpha:1});
        \UNTIL The reduction of the objective function $J$ is below the preset threshold $\varepsilon$ or the preset maximum iteration numberis reached;
        \ENSURE
			$\alpha ,\theta _n^l,y_n^{\left[ l \right]},n \in 1,2, \ldots ,N,l = 1,2.$
		\end{algorithmic}
\end{algorithm}

\section{Performance Analysis}

\subsection{Analysis of Sum-Rate}

In this subsection, we analyze the sum-rate of FILM-aided MU-MISO systems. Under ideal conditions, the FILM performs perfect ZF to yield a unit channel matrix. In practice, however, deviations arise as the FILM cannot perfectly fit the desired response, causing the resulting channel matrix to differ from the identity matrix. This relationship can be expressed as follows:
\begin{equation}
{\mathbf{G}} = \alpha {{\mathbf{I}}_K} + {\mathbf{C}},
\end{equation}
where ${\mathbf{C}} = {\left[ {\begin{array}{*{20}{c}}
  {{{\mathbf{c}}_1}}&{{{\mathbf{c}}_2}}& \cdots &{{{\mathbf{c}}_K}}
\end{array}} \right]^H}$ denotes the the channel deviation between the actual and desired channels. Consequently, the achievable rate of the $k$-th user under imperfect channel fitting can be expressed as
\begin{equation}
{R_k} = \log \left( {1 + \frac{{{\alpha ^2}{P_k}}}{{{{\left\| {{{\mathbf{c}}_k}} \right\|}^2} + \sigma _k^2}}} \right),
\end{equation}
where $P_k$ denotes the power allocated to the $k$-th stream data, and $\sigma _k^2$ represents the power of the additive white Gaussian noise (AWGN). Therefore, the sum-rate is given by
\begin{equation}\label{CC:1}
R = \sum\limits_{k = 1}^K {{R_k}}
\end{equation}

As the distribution of ${\mathbf{c}}$ is analytically intractable, a closed-form expression for (\ref{CC:1}) is hard to obtain. Nevertheless, the goal of FILM is to achieve ideal channel fitting, i.e., ${\mathbf{C}} = {\mathbf{0}}$, in which case the sum-rate reaches its upper bound
\begin{equation}
R \leqslant \sum\limits_{k = 1}^K {\log \left( {1 + \frac{{{\alpha ^2}{P_k}}}{{\sigma _k^2}}} \right).}
\end{equation}

Maximizing this upper bound can be accomplished via the classical water-filling algorithm, given by
\begin{equation}
{P_k} = \max \left( {0,\frac{1}{\lambda } - \frac{{\sigma _k^2}}{{{\alpha ^2}}}} \right),
\end{equation}
where $\lambda $ is a threshold selected to satisfy the total transmit power constraint $\sum\limits_{k = 1}^K {{P_k}}  = {P_t}$. When all users experience identical AWGN levels, i.e., $\sigma _1^2 = \sigma _2^2 =  \cdots  = \sigma _K^2 = \sigma ^2$, the optimal power allocation reduces to uniform allocation, ${P_k} = {P_t}/K$. In this case, the upper bound simplifies to
\begin{equation}\label{CC:2}
R \leqslant K\log \left( {1 + \frac{{{\alpha ^2}{P_t}}}{{K{\sigma ^2}}}} \right).
\end{equation}

\begin{remark}
Eq. (\ref{CC:2}) reveals that, under perfect ZF precoding enabled by FILM, $S$ individual sub-channels are established to support multi-stream transmission from the BS to the MUs.
\end{remark}

\subsection{Analysis of Computational Complexity}

In this subsection, we analyze the computational complexity of the proposed AO algorithm. For the proposed FILM, the computation of the optimal $\theta _n^l$ with (\ref{ps:1}) incurs a complexity of $\mathcal{O}\left( {{K^2}N} \right)$. In addition, the computations of $y_n^{\left[ 1 \right]}$ and $y_n^{\left[ 2 \right]}$ based on (\ref{fir:1}) and (\ref{sec:1}), respectively, each require a complexity of $\mathcal{O}(KN^2)$. Furthermore, the calculation of $\alpha$ via (\ref{alpha:1}) involves a complexity of $\mathcal{O}\left( {{K}} \right)$. Let $I$ denote the total number of iterations, the overall complexity is given by $\mathcal{O}\left( {{K^2}{N^2} + 2K{N^3}} \right)$. This analysis indicates that the complexity of Algorithm 1 scales cubically with respect to $N$, especially for large-scale FILMs. Notably, similar complexity analysis holds for the single-layer FIM configuration.


\section{Numerical Simulations}

In the following, we provide numerical simulations to verify the advantages of the proposed two-layer FILM-aided MU-MISO systems.

\subsection{Simulation Setups}

\begin{table}[t]
\centering
\caption{Simulation Parameters and Descriptions}
\vspace{+5mm}
\begin{tabular}{|c|c|c|}
\hline
\textbf{Parameter} & \textbf{Value} & \textbf{Description} \\
\hline
\multicolumn{3}{|c|}{\textbf{Simulation Setups}} \\
\hline
${P_t}$& 30 dBm&Transmit signal power\\
$\sigma ^2$&$-$125 dBm&Receive noise power\\
$f_0$ & 28 GHz & Carrier frequency\\
$\lambda$ & 10.7 mm & Corresponding wavelength \\
$K$&4&Number of users\\
$\left\{ {{\theta _1},{\varphi _1}} \right\}$ & $\left\{ {\frac{\pi }{6},\frac{\pi }{3}} \right\}$ & Elevation and azimuth angles for user 1\\
$\left\{ {{\theta _2},{\varphi _2}} \right\}$ & $\left\{ {\frac{\pi }{6},\frac{{2\pi }}{3}} \right\}$ & Elevation and azimuth angles for user 2\\
$\left\{ {{\theta _3},{\varphi _3}} \right\}$ & $\left\{ {\frac{\pi }{3},\frac{\pi }{4}} \right\}$ & Elevation and azimuth angles for user 3\\
$\left\{ {{\theta _4},{\varphi _4}} \right\}$ & $\left\{ {\frac{{3\pi }}{4},\frac{\pi }{2}} \right\}$ & Elevation and azimuth angles for user 4\\
$b$ & 2.5 & Path loss exponent\\
${{d_i}}$ & 150 m & Distances from FILM to users\\
$M$ & $M = K$ & Number of RF chains at the BS\\
${y^{[0]}}$ & $-$10 mm & $y$-coordinates of BS antennas\\
${x^{[0]}}$ & 0 mm & $x$-coordinates of BS antennas\\
$\left| {\Delta z_n^{\left[ 0 \right]}} \right|$ & $\lambda /2$ & BS antenna spacing along the $z$-axis\\

\hline
\multicolumn{3}{|c|}{\textbf{FILM Parameters}} \\
\hline
$L$ & 2  & Numbers of FILM layers\\
$N$ & 100  & Numbers of meta-atoms on each layer\\
$N_x$& 10 &  Number of meta-atoms along the $x$-axis\\
$N_z$& 10 &  Number of meta-atoms along the $z$-axis\\
${d_{x}}$& $\lambda /2$ & Meta-atom spacing along the $x$-axis\\
${d_{z}}$& $\lambda /2$ & Meta-atom spacing along the $z$-axis\\
${\bar y^{[2]}}$ & 0 mm & Second-layer reference along the $y$-axis\\
${\bar y^{[1]}}$ & $-$5 mm & First-layer reference along the $y$-axis\\
$\left| {{y_{\max }}} \right|$ & 2.4 mm & Maximum morphing range on $y$-axis\\
\hline
\multicolumn{3}{|c|}{\textbf{SIM Parameters}} \\
\hline
$L$ & 7 & Number of conventional SIM layers \\
$N$& 100 & Numbers of meta-atoms on each layer \\
${r_{e,t}}$& $\lambda /2$ & Adjacent meta-atom spacing\\
$d$ & 0.83 mm & Adjacent layer spacing \\
\hline
\multicolumn{3}{|c|}{\textbf{FIM Parameters}} \\
\hline
$L$ & 1 & Number of conventional FIM layers \\
$N$& 100 & Numbers of meta-atoms\\
${r_{e,t}}$& $\lambda /2$ & Adjacent meta-atom spacing\\
\hline
\multicolumn{3}{|c|}{\textbf{AO Algorithm Initialization}} \\
\hline
${{\mathbf{\Phi }}^l},l \in \mathcal{L}$ & ${{\mathbf{I}}_N}$ & Identity phase shift matrices \\
$\alpha $ & 1 & Initial channel gain \\
$I_{\max}$ & 20 & Maximum number of iterations \\
$\eta $ & 0.0001 & Step size\\
\hline
\end{tabular}
\label{tab:sim_parameters}
\end{table}

The simulation settings are summarized in Table \ref{tab:sim_parameters}. The transmit power is configured as ${P_t} = 30$ dBm, while the receive noise power at users is configured as ${N_0} =  - 125$ dBm. Moreover, a total of $K = 4$ users are considered, located at elevation/azimuth angles of $\left\{ {\frac{\pi }{6},\frac{\pi }{3}} \right\},\left\{ {\frac{\pi }{6},\frac{{2\pi }}{3}} \right\},\left\{ {\frac{\pi }{3},\frac{\pi }{4}} \right\},\left\{ {\frac{{3\pi }}{4},\frac{\pi }{2}} \right\}$, respectively. Furthermore, the carrier frequency is set to${f_0} = 28$ GHz, corresponding to a wavelength of $\lambda  = 10.7$ mm \cite{An1}. For simplicity, the path loss exponent is set to $b = 2.5$, and the distances from the second layer of FILM to each user are set to $d_i=150$ m. The number of RF chains at the BS is equal to the number of users. The BS antenna array is deployed along the $z$-axis with inter-element spacing of $\lambda /2$, located at ${y^{\left[ 0 \right]}} =  - 10$ mm, and aligned with the center of the FILM, SIM, or FIM. Quadrature phase shift keying (QPSK) modulation scheme is employed.

For the proposed two-layer FILM, each layer comprises $N=100$ meta-atoms arranged in a $10 \times 10$ grid along the $x$- and $z$- axes, with adjacent meta-atom spacing of $\lambda /2$. The distance between two layers is 5 mm, and the maximum morphing range reaches 2.4 mm. The corresponding transmission coefficient matrices are given in (\ref{wnn:1}).

A conventional 7-layer SIM counterpart is considered, with an inter-layer spacing of 0.83 mm, resulting in an overall thickness of 5 mm. Each layer comprises $N =$ 100 meta-atoms, consistent with the FILM setup. The inter-layer transmission coefficients are defined as the expression in (\ref{wnn:1}). A uniform spacing of ${r_{e,t}} = \lambda /2$ between neighboring meta-atoms is assumed, as referenced in \cite{An1}.

In the single-layer FIM counterpart, the metasurface also consists of $N =$ 100 meta-atoms. Similarly, a half-wavelength separation ${r_{e,t}} = \lambda /2$ between neighboring meta-atoms is arranged.

In Algorithm 1, all phase shift matrices are initialized as identity matrices ${{\mathbf{\Phi }}^l} = {{\mathbf{I}}_N},l \in \mathcal{L}$, and the channel gain is initialized as $\alpha = 1$. The algorithm runs for a maximum of 20 iterations, with a fixed step size of 0.0001.

To ensure a fair comparison, multiple evaluation metrics are employed. The first metric is the normalized mean square error (NMSE) between the actual and the target identity channels, expressed as
\begin{equation}
\Delta  = \mathbb{E}\left( {\frac{{\left\| {{\mathbf{HP}} - \alpha {{\mathbf{I}}_K}} \right\|_F^2}}{{\left\| {\alpha {{\mathbf{I}}_K}} \right\|_F^2}}} \right).
\end{equation}
The second metric is the sum-rate, expressed in (\ref{CC:1}). Its theoretical upper bound of the sum-rate is provided in (\ref{CC:2}). The BER is calculated as the ratio of the number of erroneously decoded bits to that of total transmitted bits.

\begin{figure}[t]
\centering
\includegraphics[width=3.5in]{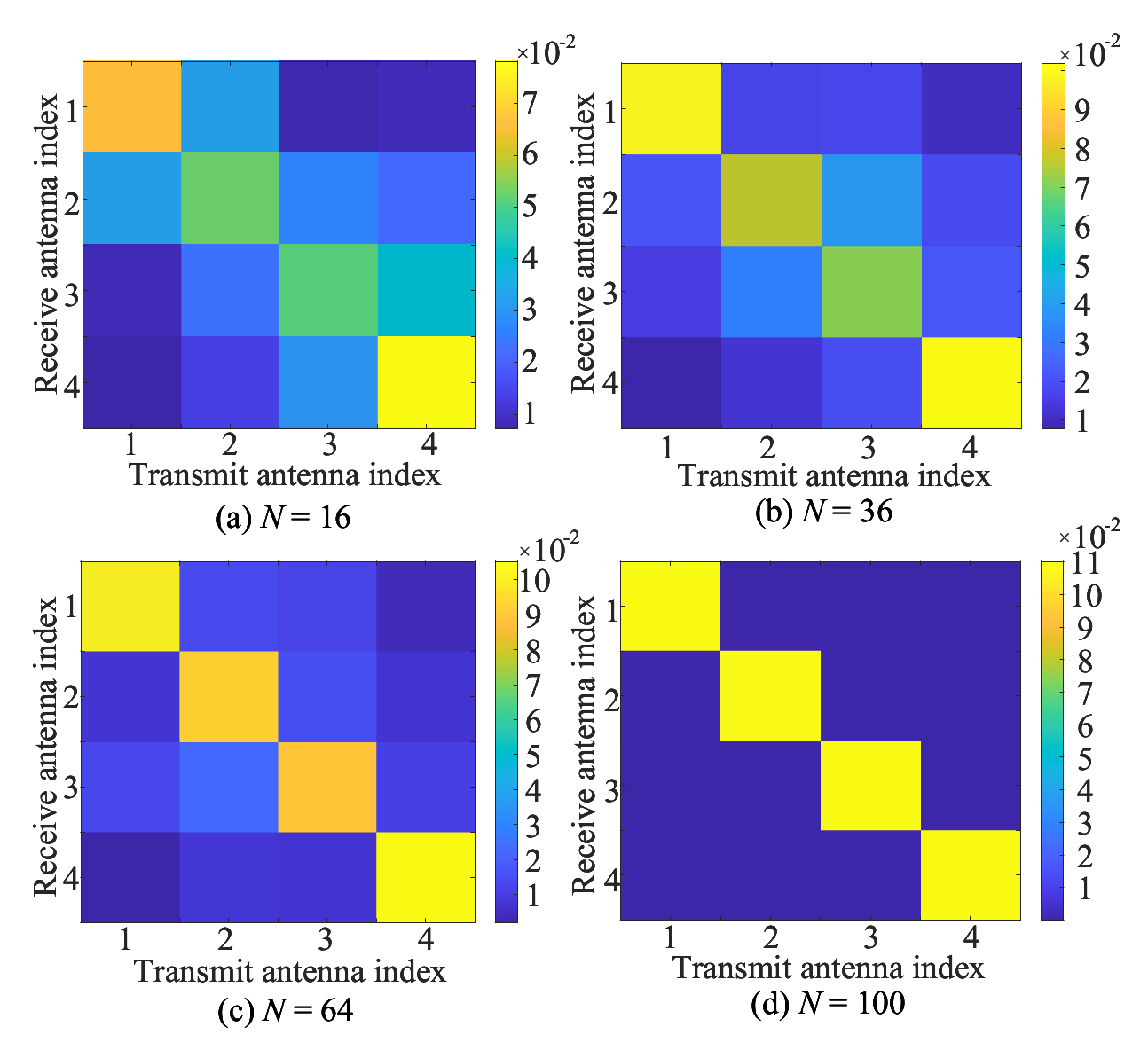}
\vspace{-3mm}
\caption{Visualization of an actual channel ${\mathbf{HP}}$.}
\label{fig_2}
\vspace{-0em}
\end{figure}
\begin{figure}[t]
\centering
\includegraphics[width=3.5in,height=2.7in]{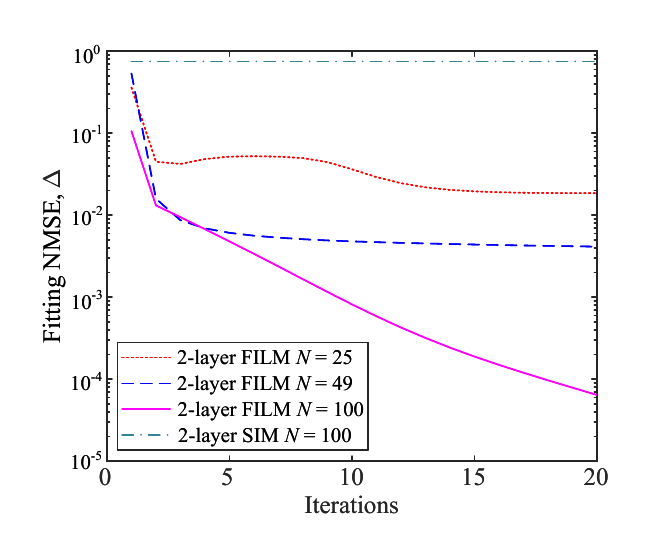}
\vspace{-3mm}
\caption{NMSE convergence behavior over iterations.}
\label{fig_3}
\vspace{-0em}
\end{figure}

\begin{figure}[t]
\centering
\includegraphics[width=3.5in,height=2.7in]{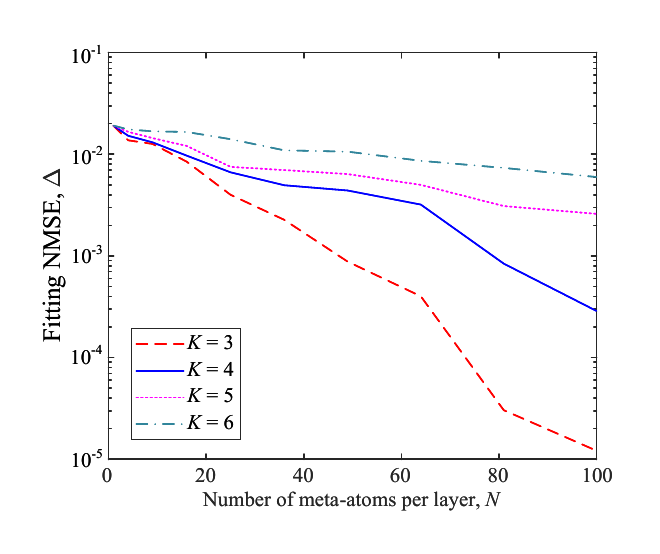}
\vspace{-3mm}
\caption{NMSE versus number of meta-atoms per layer.}
\label{fig_4}
\vspace{-0em}
\end{figure}

\begin{figure}[t]
\centering
\includegraphics[width=3.5in,height=2.7in]{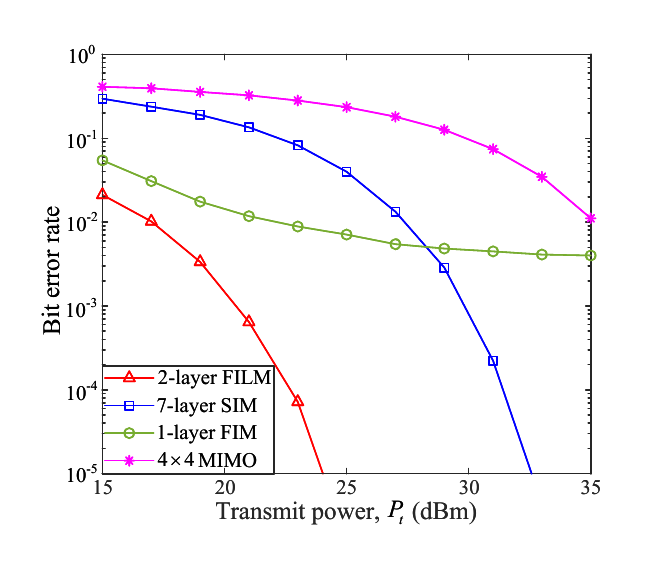}
\vspace{-3mm}
\caption{BER comparison of FILM, FIM, SIM, and MIMO versus transmit power.}
\label{fig_12}
\vspace{-0em}
\end{figure}

\begin{figure}[t]
\centering
\includegraphics[width=3.5in,height=2.7in]{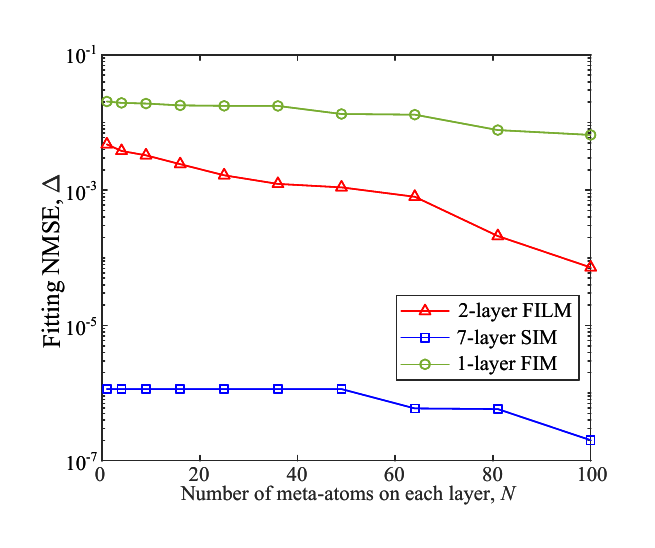}
\vspace{-3mm}
\caption{NMSE comparison of FILM, FIM, and SIM versus number of meta-atoms on each layer.}
\label{fig_8}
\vspace{-0em}
\end{figure}

\begin{figure}[t]
\centering
\includegraphics[width=3.5in,height=2.7in]{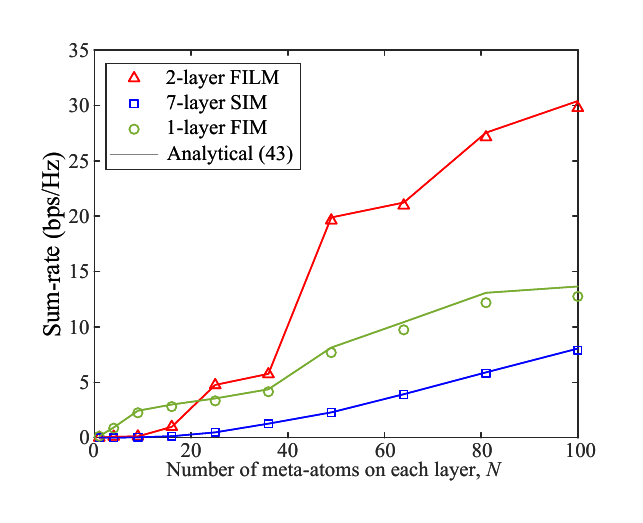}
\vspace{-3mm}
\caption{Sum-rate comparison of FILM, FIM, and SIM versus number of meta-atoms on each layer.}
\label{fig_9}
\vspace{-0em}
\end{figure}

\begin{figure}[t]
\centering
\includegraphics[width=3.5in,height=2.7in]{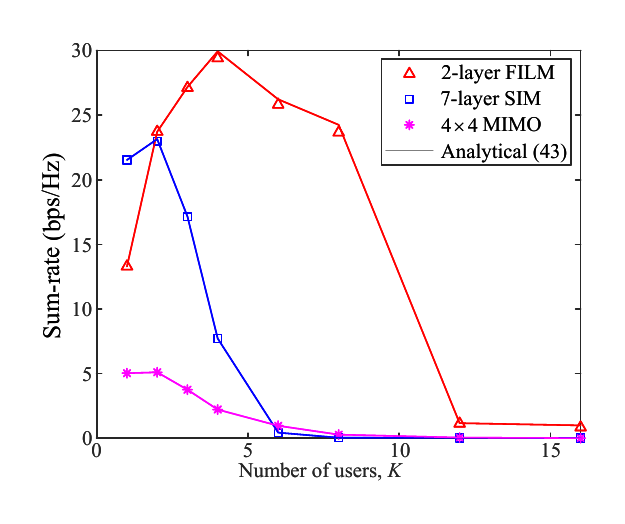}
\vspace{-3mm}
\caption{Sum-rate comparison of FILM, SIM, and MIMO versus number of users.}
\label{fig_10}
\vspace{-0em}
\end{figure}

\begin{figure}[t]
\centering
\includegraphics[width=3.5in,height=2.7in]{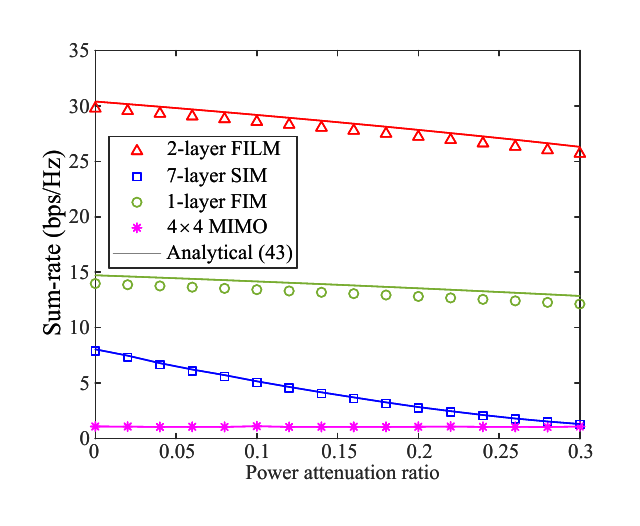}
\vspace{-3mm}
\caption{Sum-rate comparison of FILM, FIM, SIM, and MIMO versus power attenuation ratio.}
\label{fig_11}
\vspace{-0em}
\end{figure}

\begin{figure}[t]
\centering
\includegraphics[width=3.5in,height=2.7in]{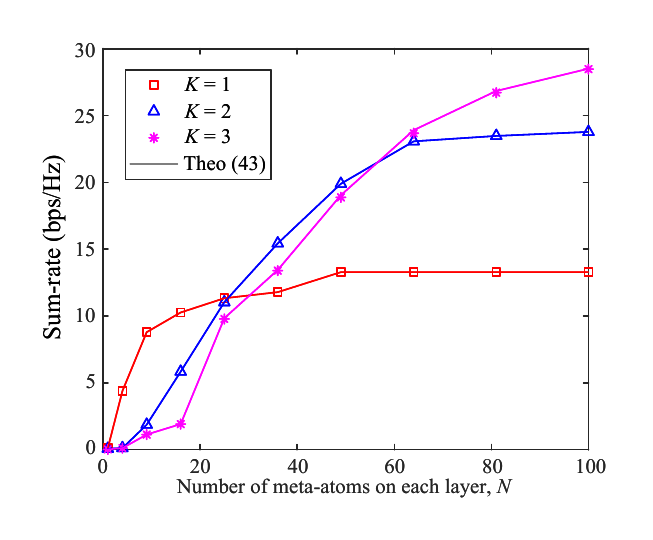}
\vspace{-3mm}
\caption{Sum-rate versus number of meta-atoms on each layer.}
\label{fig_5}
\vspace{-0em}
\end{figure}

\begin{figure}[t]
\centering
\includegraphics[width=3.5in,height=2.7in]{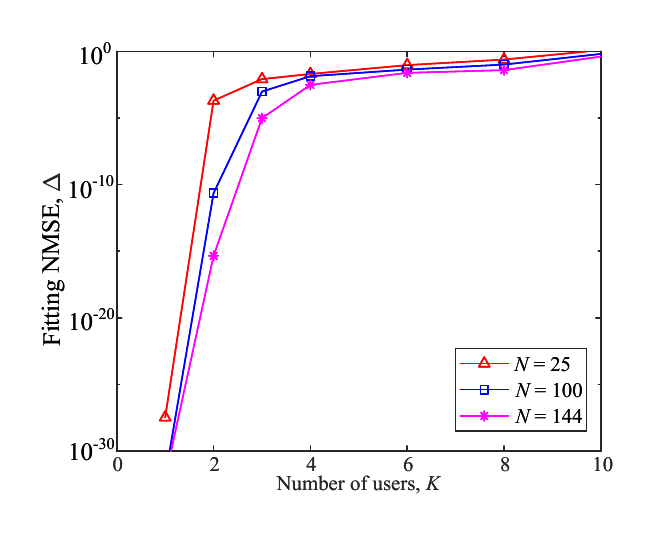}
\vspace{-3mm}
\caption{NMSE versus number of users.}
\label{fig_6}
\vspace{-0em}
\end{figure}

\begin{figure}[t]
\centering
\includegraphics[width=3.5in,height=2.7in]{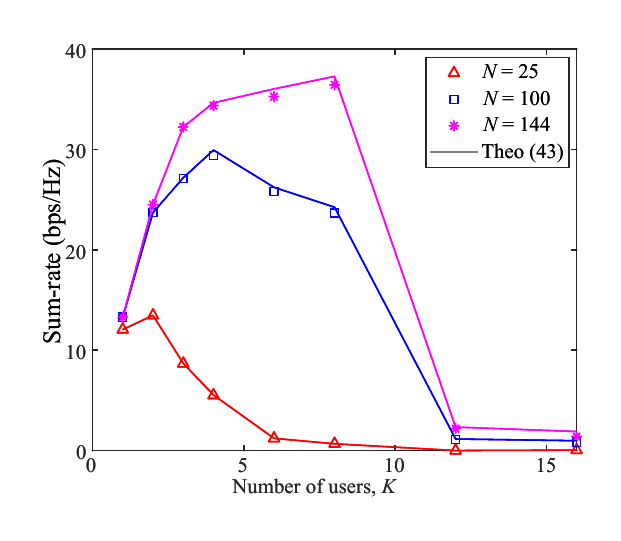}
\vspace{-3mm}
\caption{Sum-rate versus number of users.}
\label{fig_7}
\vspace{-0em}
\end{figure}

\begin{figure}[t]
\centering
\includegraphics[width=3.5in,height=2.7in]{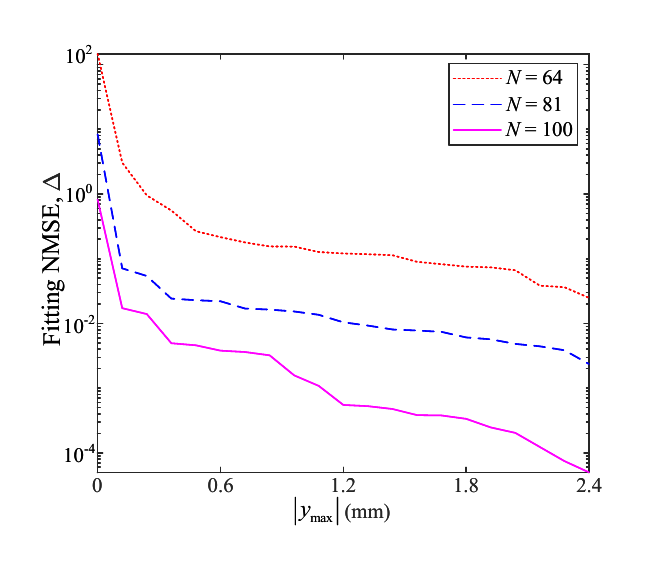}
\vspace{-3mm}
\caption{NMSE versus maximum morphing range.}
\label{fig_13}
\vspace{-0em}
\end{figure}

\begin{figure}[t]
\centering
\includegraphics[width=3.5in,height=2.7in]{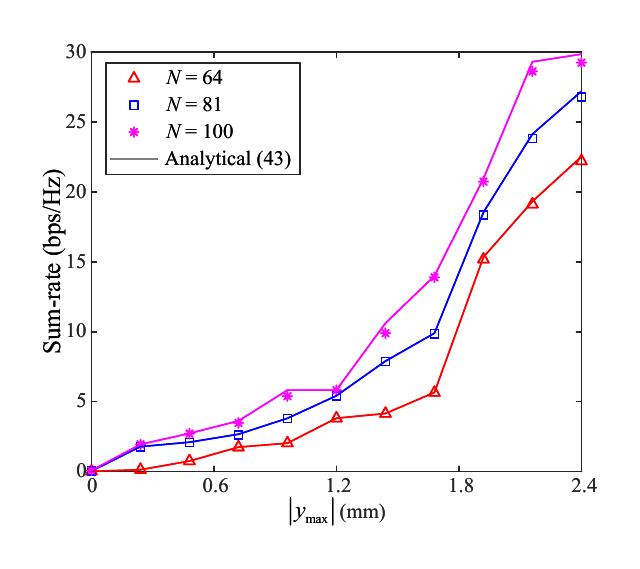}
\vspace{-3mm}
\caption{Sum-rate versus maximum morphing range.}
\label{fig_14}
\vspace{-0em}
\end{figure}

\subsection{Convergence Performance of the Proposed Algorithm}

In Fig. \ref{fig_2}, the end-to-end equivalent channel ${\mathbf{HP}}$ is depicted for various values of meta-atom count $N$. As expected, for a smaller number of meta-atoms $N = 16$, the FILM architecture fails to approximate a diagonal channel matrix. This indicates significant inter-stream interference, as the channel is unable to sufficiently decouple the transmitted data streams. For a larger number of meta-atoms $N=100$, the channel matrix becomes nearly diagonal, effectively approaching an identity matrix. This behavior reflects the ability of a two-layer FILM to construct a well-conditioned end-to-end channel, enabling each data stream to be transmitted and received independently. Consequently, the interference is mitigated, and the system design is simplified, reducing the processing complexity at both the transmitter and receiver sides.

Fig. \ref{fig_3} illustrates the NMSE convergence behavior over iterations. As the number of iterations increases, the NMSE exhibits a clear downward trend, with significantly faster convergence observed when the number of meta-atoms in the FILM structure is larger. For instance, when $N = 100$, the NMSE drops below ${10^{ - 4}}$ after 20 iterations. In contrast, the two-layer SIM baseline consistently maintains an NMSE above ${10^{ - 1}}$, indicating its limited capability in channel fitting. This comparison highlights the superiority of the proposed FILM architecture, where the deformation characteristics introduced by flexible materials effectively enhance the channel fitting performance, particularly in scenarios involving a small number of layers.

Fig. \ref{fig_4} presents the relationship between the fitting NMSE and the number of meta-atoms in each layer $N$. It can be observed that the fitting NMSE decreases monotonically with an increasing number of meta-atoms, indicating enhanced adjustment capability of the FILM due to the additional DoFs provided for channel fitting. Furthermore, for a fixed meta-atom count, increasing the number of users results in a higher fitting NMSE, suggesting that accommodating more users poses greater challenges for FILM channel fitting. Hence, balancing the number of users and fitting accuracy emerges as an important trade-off warranting further investigation.

\emph{In summary, the proposed algorithm demonstrates stable and efficient convergence. As the number of meta-atoms increases, the NMSE decreases rapidly, reaching below ${10^{ - 4}}$ within 20 iterations, outperforming conventional 2-layer SIM and 1-layer FIM. NMSE decreases monotonically with more meta-atoms but increases with the number of users, highlighting a trade-off between fitting accuracy and MU multiplexing gain.}

\subsection{Performance Comparison to Benchmark Schemes}

Fig. \ref{fig_12} compares the BER performance of the proposed FILM architecture with conventional SIM, FIM, and MIMO counterparts. As illustrated, the BER of all schemes decreases with increasing transmit power. Across the evaluated SNR range, FILM achieves the lowest BER, followed by SIM and then MIMO. While FIM exhibits BER levels comparable to FILM at low SNR, its performance saturates as the SNR increases, with the BER converging to a floor near ${10^{ - 2}}$. This plateau is attributed to limited channel fitting accuracy and substantial inter-stream interference inherent to the FIM structure. In contrast, the superior fitting capability of FILM enables it to maintain low BER across the full SNR range, highlighting its robustness and effectiveness in high-throughput, low-error communication scenarios.

Fig. \ref{fig_8} shows the NMSE comparison among the proposed FILM, the conventional single-layer FIM, and the conventional seven-layer SIM counterparts. As illustrated, the NMSE for all three schemes decreases as the per-layer meta-atom count grows. Among them, the SIM architecture consistently achieves the lowest NMSE, followed by FILM, while FIM yields the highest error across all configurations. This trend highlights the increasing difficulty of channel fitting with a reduced number of structural layers. Nevertheless, the incorporation of flexible materials significantly enhances the fitting capability. Specifically, FIM achieves an NMSE below ${10^{ - 2}}$ with $N=100$, whereas FILM further reduces the NMSE to below
${10^{ - 4}}$. These findings confirm the efficiency of the proposed FILM design, which achieves near-SIM-level fitting accuracy using only a two-layer structure, thus striking a favorable trade-off between performance and structural simplicity.

Fig. \ref{fig_9} presents a comparison of the sum-rate performance for the proposed FILM architecture against the single-layer FIM and seven-layer SIM counterparts. As observed, the sum-rate of all three schemes rises as per-layer meta-atom count grows. Notably, FILM consistently outperforms SIM, which becomes more pronounced at a larger number of meta-atoms. This is primarily due to FILM's enhanced channel fitting capability, which benefits significantly from the increased DoFs introduced by additional meta-atoms. While FIM exhibits a slight advantage under extremely low meta-atom counts, its performance plateaus as the count increases, ultimately falling between that of FILM and SIM. Furthermore, a clear discrepancy exists between FIM's actual sum-rate and its theoretical bound, indicating limitations in its ability to decouple data streams and suppress inter-stream interference. In contrast, FILM achieves superior sum-rate performance while maintaining effective channel fitting, and this advantage becomes increasingly prominent with higher meta-atom densities.

Fig. \ref{fig_10} illustrates the sum-rate performance of the proposed FILM architecture in comparison with the SIM and conventional MIMO schemes under varying numbers of users. With an increasing number of users, all three schemes exhibit a characteristic trend: the sum-rate initially improves due to enhanced spatial multiplexing but subsequently degrades as channel fitting becomes increasingly challenging. Specifically, under low user densities, the channel fitting task is relatively manageable, thus the system throughput benefits from additional data streams. However, as user density continues to rise, the degradation in channel fitting accuracy leads to increased inter-stream interference and a reduction in sum-rate. Notably, the FILM architecture outperforms the SIM counterpart, which in turn surpasses conventional MIMO. Furthermore, FILM demonstrates a higher user capacity and achieves a greater peak sum-rate, highlighting its superior scalability and channel fitting capability.

Fig. \ref{fig_11} quantifies the impact of the power attenuation ratio on the achievable sum-rate. The power attenuation ratio characterizes the signal loss incurred as it traverses each layer of metasurface structure. In ideal conditions, signal propagation through multiple layers is assumed lossless; however, in practical scenarios, a portion of the signal power is inevitably attenuated at each layer. For instance, an attenuation ratio of 0.1 implies that 90\% of the signal power is retained after passing through one layer. As observed from Fig. \ref{fig_11}, as the power attenuation ratio increases, the sum-rate degradation in the SIM architecture is significantly pronounced, whereas FILM exhibits a milder decline, slightly exceeding that of FIM. In contrast, the performance of conventional MIMO remains virtually unaffected due to the absence of metasurface layers. This sensitivity is attributed to SIM's reliance on a deeper metasurface structure, comprising seven layers, in contrast to two layers in FILM, one layer in FIM, and none in MIMO. In terms of absolute performance, FILM achieves the highest sum-rate, followed by FIM, SIM, and finally MIMO. These findings confirm that while increased attenuation degrades performance across all architectures, the proposed FILM demonstrates superior robustness and retains its performance advantage under practical, lossy conditions.

\emph{In a nutshell, the proposed FILM architecture demonstrates superior overall performance compared to conventional SIM, FIM, and MIMO counterparts across multiple evaluation metrics. While SIM achieves lower NMSE due to its deeper seven-layer structure, FILM attains comparable NMSE using only two layers, thus offering a significantly more compact and efficient design. In terms of sum-rate, FILM consistently outperforms SIM and FIM, particularly as the number of meta-atoms increases, owing to its enhanced channel fitting capability and reduced inter-stream interference. Compared to MIMO, FILM not only provides higher throughput but also supports a larger number of users with superior scalability. Moreover, FILM exhibits greater resilience to power attenuation across metasurface layers, maintaining its performance advantage under practical lossy conditions. In BER analysis, FILM achieves the lowest error rates across a wide SNR range, reinforcing its robustness and reliability. Overall, FILM strikes a balance between architectural simplicity and performance, making it a compelling solution for future high-capacity, low-error wireless systems.}

\subsection{Performance Variation versus System Parameters}

Fig. \ref{fig_5} evaluates the sum-rate as a function of the per-layer meta-atom count $N$. As observed, for a fixed number of users $K$, increasing the per-layer meta-atom count $N$ improves the sum-rate performance. Conversely, with a small $N$, a larger $K$ may reduce the sum-rate due to the greater challenge in precisely matching the desired channel. However, with sufficiently large $N$, increasing $K$ leads to a higher sum-rate as more streams can be reliably transmitted. Furthermore, the close alignment between the theoretical bound and the simulation results validates the accuracy of (\ref{CC:2}).

Figs. \ref{fig_6} and \ref{fig_7} quantify the fitting NMSE and sum-rate, respectively, versus the number of users $K$. As illustrated in Fig. \ref{fig_6}, the NMSE increases with the number of users, indicating that higher traffic loads complicate the channel fitting process. Fig. \ref{fig_7} further presents the relationship between sum-rate and the number of transmitted data streams. The system sum-rate first rises and then declines as the number of users increases. This behavior is attributed to the fact that, under a small number of users, the fitting accuracy is easily maintained, and more users raises the total data streams, thereby improving the sum-rate. However, beyond a certain point, the degradation in fitting accuracy becomes the dominant limiting factor, significantly impairing system performance and reducing the achievable sum-rate. Moreover, increasing the per-layer meta-atom count $N$ not only enhances the system's capacity to accommodate more users but also elevates the peak sum-rate. These observations highlight essential balance between channel-fitting precision and the quantity of data streams in maximizing system throughput.

Figs. \ref{fig_13} and \ref{fig_14} shows the fitting NMSE and sum-rate, respectively, versus the maximum morphing range $\left| {{y_{\max }}} \right|$. As the maximum morphing range increases, the NMSE decreases, while the sum-rate exhibits an increasing trend. However, the improvement in NMSE marginally diminishes with a larger morphing range, indicating that even a fine-tuning deformation may substantially enhance FILM's signal processing performance. When the maximum morphing range is small, the sum-rate increases gradually to preserve fitting accuracy, followed by a short phase of rapid growth before reaching saturation. Furthermore, increasing the number of units per layer in FILM tends to yield a lower NMSE and a higher sum-rate.

\emph{In general, the performance of the proposed FILM architecture is jointly determined by the number of meta-atoms per layer $N$, the number of users $K$, and the maximum morphing range $\left| {{y_{\max }}} \right|$. Increasing $N$ enhances the channel fitting capability by introducing additional DoFs, thereby reducing the NMSE, mitigating inter-stream interference, and improving the overall sum-rate. Conversely, increasing $K$ introduces fitting challenges, which may degrade NMSE and sum-rate when $N$ is insufficient. Therefore, an optimal trade-off between $N$ and $K$ is essential to maximize system throughput while maintaining accurate end-to-end channel realization. Furthermore, increasing the maximum morphing range on the wavelength scale may lead to smaller NMSE and higher sum-rate.}

\section{Conclusion}

In this work, a novel FILM architecture was proposed to enhance power efficiency and signal processing capabilities in MU-MISO systems. Specifically, a two-layer FILM-assisted MU-MISO framework was developed, wherein the channel fitting problem was addressed via joint optimization of the FILM shape and phase shifts. In order to reduce the computational complexity, an efficient AO algorithm was proposed, incorporating closed-form updates for phase shifts and a GD-based method for shape optimization. Theoretical upper bounds on sum-rate were derived, and the computational complexity of the proposed algorithm was analyzed. Numerical results demonstrated that the proposed FILM architecture achieves over 200\% improvement in sum-rate and more than 7 dB BER gain compared to the conventional seven-layer SIM structure.

\setcounter{subsubsection}{0}

\ifCLASSOPTIONcaptionsoff
  \newpage
\fi


\begin{thebibliography}{00}
\bibitem{SIM01}
M. Di Renzo, ``State of the art on stacked intelligent metasurfaces: Communication, sensing and computing in the wave domain,'' in \emph{19th Eur. Conf. Antennas Propag. (EuCAP)}, Stockholm, Sweden, Apr. 2025, pp. 1-3.
\bibitem{SIM02}
X. Yao, J. An, L. Gan, M. Di Renzo, and C. Yuen, ``Channel estimation for stacked intelligent metasurface-assisted wireless networks,'' \emph{IEEE Wireless Commun. Lett.}, vol. 13, no. 5, pp. 1349-1353, May 2024.
\bibitem{SIM03}
J. An, M. Di Renzo, M. Debbah, and C. Yuen, ``Stacked intelligent metasurfaces for multiuser beamforming in the wave domain,'' in \emph{Proc. IEEE Int. Conf. Commun.}, Rome, Italy, May 2023, pp. 2834-2839.
\bibitem{SIM04}
N. Chamanara, Y. Vahabzadeh, and C. Caloz, ``Stacked metasurface slab,'' in \emph{2018 12th Int. Congr. Artif. Mater. Novel Wave Phenomena (Metamaterials)}, Espoo, Finland, Aug. 2018, pp. 70-72.
\bibitem{SIM05}
N. U. Hassan, J. An, M. Di Renzo, M. Debbah, and C. Yuen, ``Efficient beamforming and radiation pattern control using stacked intelligent metasurfaces,'' \emph{IEEE Open J. Commun. Soc.}, vol. 5, pp. 599-611, Jan. 2024.
\bibitem{RIS01}
Y. Liu, X. Mu, X. Liu, M. Di Renzo, Z. Ding, and R. Schober, ``Reconfigurable intelligent surface-aided multi-user networks: Interplay between NOMA and RIS,'' \emph{IEEE Wireless Commun.}, vol. 29, no. 2, pp. 169-176, Apr. 2022.
\bibitem{RIS02}
M. Di Renzo, F. H. Danufane, and S. Tretyakov, ``Communication models for reconfigurable intelligent surfaces: From surface electromagnetics to wireless networks optimization,'' \emph{Proc. IEEE}, vol. 110, no. 9, pp. 1164-1209, Sep. 2022.
\bibitem{RIS03}
X. Mu, J. Xu, Y. Liu, and L. Hanzo, ``Reconfigurable intelligent surface-aided near-field communications for 6G: Opportunities and challenges,'' \emph{IEEE Veh. Technol. Mag.}, vol. 19, no. 1, pp. 65-74, Mar. 2024.
\bibitem{RIS04}
Y. Liu \emph{et al.}, ``Reconfigurable Intelligent Surfaces: Principles and opportunities,'' \emph{IEEE Commun. Surv. \& Tutor.}, vol. 23, no. 3, pp. 1546-1577, 3rd Quart. 2021.
\bibitem{RIS05}
T. Ma, Y. Xiao, X. Lei, P. Yang, X. Lei and O. A. Dobre, ``Large intelligent surface assisted wireless communications with spatial modulation and antenna selection,'' \emph{IEEE J. Sel. Areas Commun.}, vol. 38, no. 11, pp. 2562-2574, Nov. 2020.
\bibitem{RIS06}
T. Wu \emph{et al.}, ``Exploit high-dimensional RIS information to localization: What is the impact of faulty element?'' \emph{IEEE J. Sel. Areas Commun.}, vol. 42, no. 10, pp. 2803-2819, Oct. 2024.
\bibitem{RIS07}
E. Shi \emph{et al.}, ``RIS-aided cell-free massive MIMO systems for 6G: Fundamentals, system design, and applications,'' \emph{Proc. IEEE}, vol. 112, no. 4, pp. 331-364, Apr. 2024.
\bibitem{RIS08}
H. Niu, X. Lei, Y. Xiao, M. Xiao, and S. Mumtaz, ``On the efficient design of RIS-assisted secure MISO transmission,'' \emph{IEEE Wireless Commun. Lett.}, vol. 11, no. 8, pp. 1664-1668, Aug. 2022.
\bibitem{RIS09}
C. Huang, A. Zappone, G. C. Alexandropoulos, M. Debbah, and C. Yuen, ``Reconfigurable intelligent surfaces for energy efficiency in wireless communication,'' \emph{IEEE Trans. Wireless Commun.}, vol. 18, no. 8, pp. 4157-4170, Aug. 2019.
\bibitem{RIS10}
H. Niu, Y. Xiao, X. Lei, L. Dan, W. Xiang, and C. Yuen, ``Reconfigurable intelligent surface-assisted passive beamforming attack,'' \emph{IEEE Trans. Inf. Forensics Secur.}, vol. 19, pp. 8236-8247, 2024.

\bibitem{SIM06}
J. An \emph{et al.}, ``Stacked intelligent metasurface-aided MIMO transceiver design,'' \emph{IEEE Wireless Commun.}, vol. 31, no. 4, pp. 123-131, Aug. 2024.
\bibitem{SIM07}
Y. Sun \emph{et al.}, ``Active-passive cascaded RIS-aided receiver design for jamming nulling and signal enhancing,'' \emph{IEEE Trans. Wireless Commun.}, vol. 23, no. 6, pp. 5345-5362, Jun. 2024.
\bibitem{SIM08}
K. An \emph{et al.}, ``Exploiting multi-layer refracting RIS-assisted receiver for HAP-SWIPT networks,'' \emph{IEEE Trans. Wireless Commun.}, vol. 23, no. 10, pp. 12638-12657, Oct. 2024.
\bibitem{SIM10}
J. An \emph{et al.}, ``Two-dimensional direction-of-arrival estimation using stacked intelligent metasurfaces,'' \emph{IEEE J. Sel. Areas Commun.}, vol. 42, no. 10, pp. 2786-2802, Oct. 2024.
\bibitem{beam1}
M. Nerini and B. Clerckx, ``Physically consistent modeling of stacked intelligent metasurfaces implemented with beyond diagonal RIS,'' \emph{IEEE Commun. Lett.}, vol. 28, no. 7, pp. 1693-1697, Jul. 2024.
\bibitem{beam2}
H. Liu, \emph{et al.}, ``Stacked intelligent metasurfaces for wireless sensing and communication: Applications and challenges.'' arXiv: 2407.03566 [cs.IT], Jul. 2024.
\bibitem{beam3}
A. Papazafeiropoulos, \emph{et al.}, ``Achievable rate optimization for large stacked intelligent metasurfaces based on statistical CSI,'' \emph{IEEE Wireless Commun. Lett.}, vol. 13, no. 9, pp. 2337-2341, Sep. 2024.
\bibitem{focus1}
H. Niu, J. An, A. Papazafeiropoulos, L. Gan, S. Chatzinotas, and M. Debbah, ``Stacked intelligent metasurfaces for integrated sensing and communications,'' \emph{IEEE Wireless Commun. Lett.}, vol. 13, no. 10, pp. 2807-2811, Oct. 2024.
\bibitem{focus2}
S. Lin, J. An, L. Gan, M. Debbah, and C. Yuen, ``Stacked intelligent metasurface enabled LEO satellite communications relying on statistical CSI,'' \emph{IEEE Wireless Commun. Lett.}, vol. 13, no. 5, pp. 1295-1299, May 2024.
\bibitem{MU1}
H. Liu, J. An, G. C. Alexandropoulos, D. W. K. Ng, C. Yuen, and L. Gan, ``Multi-User MISO with stacked intelligent metasurfaces: A DRL-based sum-rate optimization approach,'' \emph{IEEE Trans. Cogn. Commun. Netw.}, Apr. 2025, (early access) doi: 10.1109/TCCN.2025.3558008.
\bibitem{MU2}
Z. Wang \emph{et al.}, ``Multi-user ISAC through stacked intelligent metasurfaces: New algorithms and experiments,'' in \emph{ Proc. IEEE Global Commun. Conf. (GLOBECOM)}, Cape Town, South Africa, Dec. 2024, pp. 4442-4447.

\bibitem{MIMO1}
E. Shi \emph{et al.}, ``Joint AP-UE association and precoding for SIM-aided cell-free massive MIMO systems,'' \emph{IEEE Trans. Wireless Commun.}, Mar. 2025, early access, doi: 10.1109/TWC.2025.3546927.
\bibitem{MIMO2}
E. Shi, J. Zhang, Y. Zhu, J. An, C. Yuen, and B. Ai, ``Uplink performance of stacked intelligent metasurface-enhanced cell-free massive MIMO systems,'' \emph{IEEE Trans. Wireless Commun.}, vol. 24, no. 5, 3731-3746, May 2025.
\bibitem{MIMO3}
Q. Li, M. El-Hajjar, C. Xu, J. An, C. Yuen, and L. Hanzo, ``Stacked intelligent metasurfaces for holographic MIMO-aided cell-free networks,'' \emph{IEEE Trans. Commun.}, vol. 72, no. 11, pp. 7139-7151, Nov. 2024.
\bibitem{secure1}
H. Niu, J. An, L. Zhang, X. Lei, and C. Yuen, ``Enhancing physical layer security for SISO systems using stacked intelligent metasurfaces,'' in \emph{2024 IEEE VTS Asia Pacific Wireless Commun. Symp. (APWCS)}, Singapore, Aug. 2024, pp. 1-5.
\bibitem{secure2}
H. Niu, X. Lei, J. An, L. Zhang, and C. Yuen, ``On the efficient design of stacked intelligent metasurfaces for secure SISO transmission,'' \emph{IEEE Trans. Inf. Forensics Secur.}, vol. 20, pp. 60-70, Nov. 2024.
\bibitem{secure3}
J. Chen \emph{et al.}, ``A survey on directional modulation: Opportunities, challenges, recent advances, implementations, and future trends,'' \emph{IEEE Internet Things J.}, vol. 12, no. 16, pp. 32581-32615, 15 Aug.15, 2025.
\bibitem{secure4}
H. Niu \emph{et al.}, ``A Survey on Artificial Noise for Physical Layer Security: Opportunities, Technologies, Guidelines, Advances, and Trends, '' \emph{IEEE Commun. Surv. Tutor. }, 2025, early access, doi: 10.1109/COMST.2025.3610758.
\bibitem{Liu1}
C. Liu \emph{et al.}, ``A programmable diffractive deep neural network based on a digital-coding metasurface array,'' \emph{Nature Electron.}, vol. 5, no. 2, pp. 113-122, Feb. 2022.
\bibitem{An1}
J. An \emph{et al.}, ``Stacked intelligent metasurfaces for efficient holographic MIMO communications in 6G,'' \emph{IEEE J. Sel. Areas Commun.}, vol. 41, no. 8, pp. 2380-2396, Aug. 2023.
\bibitem{Att1}
H. Xu, \emph{et al.}, ``Multifunctional metasurfaces: Design principles and device realizations,'' \emph{Morgan $\& $ Claypool Publishers}, 2021.
\bibitem{fiber1}
H. Niu, \emph{et al.}, ``Introducing meta-fiber into stacked intelligent metasurfaces for MIMO communications: A low-complexity design with only two layers,'' \emph{IEEE Trans. Wireless Commun.}, 2025, early access, doi: 10.1109/TWC.2025.3600915.
\bibitem{FIM1}
J. An, M. Debbah, T. J. Cui, Z. N. Chen, and C. Yuen, ``Emerging technologies in intelligent metasurfaces: Shaping the future of wireless communications,'' \emph{IEEE Trans. Antennas Propag.}, 2025, early access, doi: 10.1109/TAP.2025.3571069.
\bibitem{FIM2}
J. An, C. Yuen, M. Di Renzo, M. Debbah, H. V. Poor, and L. Hanzo, ``Downlink multiuser communications relying on flexible intelligent metasurfaces,'' in \emph{Proc. IEEE Global Commun. Conf. (GLOBECOM)}, Cape Town, South Africa, Dec. 2024, pp. 4932-4937.
\bibitem{FIM3}
J. An, C. Yuen, M. D. Renzo, M. Debbah, H. V. Poor and L. Hanzo, ``Flexible intelligent metasurfaces for downlink multiuser MISO communications,'' \emph{IEEE Trans. Wireless Commun.}, vol. 24, no. 4, pp. 2940-2955, Apr. 2025.
\bibitem{FIM4}
Z. Teng, J. An, L. Gan, N. Al-Dhahir, and Z. Han, ``Flexible intelligent metasurface for enhancing multi-target wireless sensing,'' \emph{IEEE Trans. Veh. Technol.}, 2025, early access, doi: 10.1109/TVT.2025.3584865.
\bibitem{FIM5}
J. An, Z. Han, D. Niyato, M. Debbah, C. Yuen and L. Hanzo, ``Flexible intelligent metasurfaces for enhancing MIMO communications,'' \emph{IEEE Trans. Commun.}, 2025, early access, doi: 10.1109/TCOMM.2025.3550318.
\bibitem{FAS1}
T. Wu, \emph{et al.}, ``Fluid antenna systems enabling 6G: Principles, applications, and research directions,'' arXiv: 2412.03839[eess.SP], Dec. 2024.
\bibitem{FAS2}
J. Yao, T. Wu, L. Zhou, M. Jin, C. Huang, and C. Yuen, ``FAS vs. ARIS: Which is more important for FAS-ARIS communication systems?,'' \emph{IIEEE Trans. Wireless Commun.}, 2025, early access, doi: 10.1109/TWC.2025.3594617.
\bibitem{MAS1}
Z. Li \emph{et al.}, ``Movable antennas enabled ISAC systems: Fundamentals, opportunities, and future directions,'' \emph{IEEE Wireless Commun.}, 2025, early access, doi: 10.1109/MWC.002.2400522.
\bibitem{w1}
X. Lin \emph{et al.}, ``All-optical machine learning using diffractive deep neural networks,'' \emph{Sci.}, vol. 361, no. 6406, pp. 1004-1008, Sep. 2018.


\end{thebibliography}
\end{document}